\documentclass[showpacs,aps,epsfig,nofootinbib]{revtex4}
\usepackage{mathrsfs}

\usepackage{graphicx}
\usepackage{amsfonts}
\usepackage{epstopdf}
\usepackage{latexsym}
\usepackage{amssymb}
\usepackage{amssymb}

\usepackage[center]{subfigure}

\begin{document}

 \newcommand{\bq}{\begin{equation}}
 \newcommand{\eq}{\end{equation}}
 \newcommand{\bqn}{\begin{eqnarray}}
 \newcommand{\eqn}{\end{eqnarray}}
 \newcommand{\nb}{\nonumber}
 \newcommand{\lb}{\label}
\newcommand{\PRL}{Phys. Rev. Lett.}
\newcommand{\PL}{Phys. Lett.}
\newcommand{\PR}{Phys. Rev.}
\newcommand{\CQG}{Class. Quantum Grav.}

\title{Dirac quasinormal modes in spherically symmetric regular black holes}

\author{Jin Li}
\email{cqstarv@hotmail.com}
\affiliation {Department of Physics, Chongqing University,
Chongqing 401331, China}
\author{Hong Ma}
\email{michaelmahong@126.com}
\affiliation {Department of Physics, Chongqing University,
Chongqing 401331, China}
\author{Kai Lin}
\email{lk314159@hotmail.com}
\affiliation {Instituto de F\'isica, Universidade de S\~ao Paulo, CP 66318, 05315-970, S\~ao Paulo, Brazil}

\date{\today}

\begin{abstract}
Using the WKB approximation, massless and massive Dirac
quasinormal modes (QNMs) are studied in spherically symmetric regular spacetimes. We
analyze the relationships between QNM frequencies and the parameters
(angular momentum number $l$, magnetic monopole charge $\beta$ and
the mass of the field $m$), and discuss the extreme charge of magnetic monopole $\beta_{e}$ for  spherically symmetric regular black holes (BHs). Furthermore, we apply an expansion method
to expand QNMs in inverse powers of $L=l+1/2$, and confirm  good
precision with $l>n$. Finally, we improve traditional finite difference method to be available in massive Dirac case, and
illuminate the dynamical evolution of massive Dirac field.

\end{abstract}


\maketitle

\section{introduction}
In General Relativity, it is important to understand how to avoid singularities in black hole spacetimes. In 1968, Bardeen firstly proposed a ``regular'' black hole, without a singularity that satisfied the weak energy condition and had metric components that fell off appropriately at large distances~\cite{RegularMetric1}. It can be interpreted as the solution for a nonlinear magnetic monopole with mass $M$ and charge $\beta$~\cite{RegularMetric}. After several years, Bronnikov, Ayon-Beato and Garcia proposed a new nonlinear electrodynamics which, when coupled to gravity, produces an exact nonsingular black hole solution that also satisfies the weak energy condition~\cite{new10,new11,RegularMetric5}. Subsequently, further analyses of singularity avoidance have been proposed in the literature. For example, Hayward obtained a simple regular black hole by requiring center flatness. In this model, $\beta$ has a relationship with the cosmological constant, $\Lambda$, such that $\beta^2 = 3M/\Lambda$~\cite{RegularMetric2}. Other BH solutions through introducing the Lagrangian for non-linear electrodynamics to first order~\cite{new8,RegularMetric3}. Dymnikova put forward an exact, regular spherically symmetric, charged solution with a de Sitter center~\cite{RegularMetric4}. All of these solutions can be described with a metric of the form
\begin{equation}
ds^{2}=-f(r)dt^{2}+f(r)^{-1}dr^{2}+r^{2}d\theta^{2}+r^{2}\sin^{2}\theta
d\phi^{2},\label{eq:metric}
\end{equation}
where specific choices of $f(r)$ distinguish between the different spacetimes. These choices of $f(r)$ are shown in Table~\ref{tabfour}.

Perturbing these space times may give a glimpse into the interior region of black holes~\cite{new9}. In order to discover the physical features of spherically symmetric regular BHs, we need to study the quasinormal modes (QNMs) generated by these perturbations. Quasinormal ringing produces several complex frequencies. The real part of the frequency corresponds to the oscillation rate and the magnitude of the imaginary part corresponds to the damping rate. Flachi and Lemos studied the quasinormal modes (QNMs) generated by scalar field perturbations and found the difference between regular black holes and the one from the ordinary ones. They indicated the real part of the QNMs' frequency followed a behavior analogous to the Reissner-Nordstr\"{o}m case but in the strong charged region the exponential damping (i.e., the imaginary part)of these modes occurred less prominently than in the Reissner-Nordstr\"{o}m case~\cite{RegularBH}. This ringing signal is a promising target of gravitational wave antennas, which would give clues to the physical properties of regular BHs. In addition to the black hole geometry, the properties of the QNM spectrum also depend on the properties of the field (e.g., spin). Even after years of study, most of the research into QNMs are for scalar, electromagnetic, and gravitational perturbations (i.e, fields with integer spins)~\cite{RegularBH,new7}. Cho has looked at Dirac QNMs in Schwarzschild BHs~\cite{Diracfield}, but in general cases, QNMs in spherically symmetric regular BHs remains unstudied. Therefore, we focus on QNMs of Dirac perturbations in these regular spacetimes.

Quasinormal modes can be described in terms of the angular momentum quantum numbers $l$ and the overtone $n$. There are a large number of numerical methods used to compute the QNMs (e.g., time domain methods~\cite{QNMmethod1, QNMmethod2}, direct integration in the frequency domain~\cite{QNMmethod3,QNMmethod31,QNMmethod32},inverse potential methods~\cite{QNMmethodjia1,QNMmethodjia2}, WKB~\cite{QNMmethod4,QNMmethod5,QNMmethod6,QNMmethod61,QNMmethod7,QNMmethod8}, etc.). Among these approaches, the WKB scheme has been shown to be more accurate for both the real and imaginary parts of the dominant QNMs with $n\leq l$~\cite{Diracfield}. Flachi and Lemos have used WKB approach to study QNMs of neutral and charged scalar field perturbations for all the regular black holes in Table I, and found the relationship between the frequencies of QNMs and parameters (such as $l$, $n$ and $\beta$). Especially they illuminated how the order of WKB approximation impacted on the frequencies of QNMs in $f_{4}(r)$~\cite{RegularBH}. Therefore, we will calculate the QNMs for massive and massless Dirac fields using the WKB approximation. Cardoso \emph{et al}. have studied null geodesics and the eikonal limit and adopted an expansion method to calculate QNMs of spherically symmetric black holes~\cite{eikonal1}, which was developed by Dolan and Ottewill in 2009~\cite{eikonal2}. In the next three years, Dolan extended the method to treat an axisymmetric system~\cite{eikonal3} and evaluated the spectra of a rotating black hole analogue~\cite{eikonal4}. The precision of this method is good enough for the case where $l \gg n$. We compare the WKB and expansion methods for the massless Dirac field. For the massive case, we improve the finite difference method to find the relationship between radial wave functions and time. We determine the dynamical evolution of the Dirac field in the time domain by the finite difference method.

The paper is organized as follows. In section~\ref{sec:regDirac}, we describe the Dirac equation in regular spacetimes. In section~\ref{sec:extremeCharge}, we investigate the extreme conditions for spherically symmetric regular black holes. Quasinormal modes for massless fields are evaluated using $3^{\rm rd}$-order WKB method and the expansion method in section~\ref{sec:masslessDirac}. The massive case is analyzed in section~\ref{sec:massiveDirac}. Conclusions and future work are presented in section~\ref{sec:conclusions}.

\begin{table}
\caption{Summary of the spherically symmetric regular black holes in the paper. All the
details can be found in the original references~\cite{RegularMetric1,RegularMetric,RegularMetric5,new10, new11,RegularMetric2,new8,RegularMetric3,RegularMetric4} listed in the
first paragraph.} \label{tabfour}

\begin{tabular}{ll}
\hline\noalign{\smallskip}
Originator of $BH$   & $f(r)$ \\
\noalign{\smallskip}\noalign{\smallskip}
Bardeen & $f_{1}(r)=1-\frac{2Mr^{2}}{(r^{2}+\beta^{2})^{3/2}}$\\
Hayward & $f_{2}(r)=1-\frac{2Mr^{2}}{r^{3}+2\beta^{2}}$\\
Ayon-Beato and
Garcia, Berej et al. & $f_{3}(r)=1-\frac{2M}{r}(1-\tanh\frac{\beta^{2}}{2Mr})$\\
Dymnikova & $f_{4}(r)=1-\frac{4M}{\pi
r}(\tan^{-1}\frac{r}{r_{0}}-\frac{rr_{0}}{r^{2}+r_{0}^{2}})$
~\text{where}~$r_{0}=\pi\beta^{2}/8M$\\
Bronnikov, Ayon-Beato and
Garcia & $f_{5}(r)=1-\frac{2Mr^{2}}{(r^{2}+\beta^{2})^{3/2}}+\frac{\beta^{2}r^{2}}{(r^{2}+\beta^{2})^{2}}$\\
\noalign{\smallskip}\hline
\end{tabular}

\end{table}

\section{Dirac equation in the regular spacetime}\label{sec:regDirac}

The general equation for Dirac perturbations with mass $m$ can be written as~\cite{Diracfieldjia,Diracfield}
\begin{equation}
[\gamma^{a}e_{a}^{\mu}(\partial_{\mu}+\Gamma_{\mu})+m]\Psi=0,
\end{equation}
where $\Gamma_{\mu}=\frac{1}{8}[\gamma^{a},\gamma^{b}]e_{a}^{\nu}e_{b\nu;\mu}$
is the spin connection, $\gamma^{a}$ are the Dirac metrices and
$e_{b\nu;\mu}=\partial_{\mu}e_{b\nu}-\Gamma_{\mu\nu}^{\alpha}e_{b\alpha}$.

We consider spherically symmetric metrics of the form given by Eq.(~\ref{eq:metric}), where $f(r)$ represents the regular spacetime function $f_i(r)$ listed in Table~\ref{tabfour}.Therefore the $e_{\nu}^{a}$ can be taken to be
\begin{equation}
e_{\nu}^{a}={\rm diag}(f(r)^{1/2},f(r)^{-1/2},r,r\sin \theta).
\end{equation}

Defining the wave function as
\begin{equation}
\Psi=f(r)^{-1/4}\Phi          ,
\end{equation}
the wave equation can be simplified to
\begin{equation}
\left[\gamma^{0}f(r)^{-1/2}\frac{\partial}{\partial t}+\gamma^{1}f(r)^{1/2}(\frac{\partial}{\partial r}+\frac{1}{r})+\gamma^{2}\frac{1}{r}(\frac{\partial}{\partial \theta}+\frac{1}{2}\cot{\theta})+\gamma^{3}\frac{1}{r\sin{\theta}}\frac{\partial}{\partial \phi}+m\right]\Phi=0.
\end{equation}
Since there are two different spin magnetic quantum numbers for
Dirac particles, it is necessary to define the wave function separately.
We use the ansatz
\begin{equation}
\Phi=  \left(
                \begin{array}{c}
                  \frac{iG^{\pm}(r)}{r}\varphi^{\pm}_{jm}(\theta,\phi) \\
                  \frac{F^{\pm}(r)}{r}\varphi^{\mp}_{jm}(\theta,\phi) \\
                \end{array}\right)
              e^{-i\omega t},
\end{equation}
and
\begin{equation}
\varphi^{+}_{jm}=  \left(
                \begin{array}{c}
                  \sqrt{\frac{l+1/2+m}{2l+1}}Y^{m-1/2}_{l} \\
                   \sqrt{\frac{l+1/2-m}{2l+1}}Y^{m+1/2}_{l} \\
                \end{array}
             \right)
~~~(\text{for}~j=l+\frac{1}{2}),
\end{equation}
\begin{equation}
\varphi^{-}_{jm}=  \left(
                \begin{array}{c}
                  \sqrt{\frac{l+1/2-m}{2l+1}}Y^{m-1/2}_{l} \\
                   -\sqrt{\frac{l+1/2+m}{2l+1}}Y^{m+1/2}_{l} \\
                \end{array}
             \right)
~~~(\text{for}~j=l-\frac{1}{2}).
\end{equation}

Since we are dealing with spherically symmetric black holes, we only concern ourselves with the radial functions ($G^{\pm}$ and $F^{\pm}$). Putting these into the Dirac equation, the radial functions satisfy
\begin{equation}
\label{eqfg}
\frac{d}{dr_{*}}\left(
                        \begin{array}{c}
                          F^{\pm} \\
                         G^{\pm} \\
                        \end{array}
                      \right)-\sqrt{f(r)}\left(
                                    \begin{array}{cc}
                                      k_{\pm}/r & m \\
                                      m & -k_{\pm}/r \\
                                    \end{array}
                                  \right)\left(
                                           \begin{array}{c}
                                             F^{\pm} \\
                                             G^{\pm} \\
                                           \end{array}
                                         \right)=\left(
                                                   \begin{array}{cc}
                                                     0 & -\omega \\
                                                     \omega & 0 \\
                                                   \end{array}
                                                 \right)\left(
                        \begin{array}{c}
                          F^{\pm} \\
                         G^{\pm} \\
                        \end{array}
                      \right),
\end{equation}
Here $d/dr_{*}=f(r)d/dr$. Making some changes of $F^{\pm}$ and $G^{\pm}$ as~\cite{Diracfield},
\begin{equation}
\left(
                        \begin{array}{c}
                          \hat{F}^{\pm} \\
                         \hat{G}^{\pm} \\
                        \end{array}
                      \right)=\left(\begin{array}{cc}
                                                     \sin\frac{\theta}{2} & \cos\frac{\theta}{2} \\
                                                     \cos\frac{\theta}{2} & -\sin\frac{\theta}{2} \\
                                                   \end{array}
                                                 \right)\left(\begin{array}{c}
                          F^{\pm} \\
                          G^{\pm} \\
                        \end{array}\right),
\end{equation}
where $\theta=\tan^{-1}(mr/|k|)$. Introducing $\hat{r}_{*}=r_{*}+\tan^{-1}(mr/|k|)/2\omega$, the decoupled equations are given as
\begin{equation}
\label{radialf1}
\frac{d\hat{F}^{\pm}}{d\hat{r}_{*}}-W_{\pm}\hat{F}^{\pm}=\omega\hat{G}^{\pm},
\end{equation}
\begin{equation}
\label{radialf2}
\frac{d\hat{G}^{\pm}}{d\hat{r}_{*}}+W_{\pm}\hat{G}^{\pm}=-\omega\hat{F}^{\pm},
\end{equation}
where
\begin{equation}
\frac{d}{d\hat{r}_{*}}=\frac{f(r)}{1+\frac{m|k_{\pm}|f(r)}{2\omega(k^{2}+m^{2}r^{2})}}\frac{d}{dr}.
\end{equation}
and
\begin{equation}
W_{\pm}=\frac{\sqrt{f(r)(\frac{k_{\pm}^{2}}{r^{2}}+m^{2})}}{1+\frac{f(r)m|k_{\pm}|}{2\omega(k_{\pm}^{2}+m^{2}r^{2})}}.
\end{equation}
Then do $d/d\hat{r}_{*}$ to Eq.(\ref{radialf1})-Eq.(\ref{radialf2}), the following derivation radial functions can be gotten
\begin{equation}
(-\frac{d^{2}}{d\hat{r}^{2}_{*}}+\bar{V}_{\pm})\hat{F}^{\pm}=\omega^{2}\hat{F}^{\pm},
\end{equation}
\begin{equation}
(-\frac{d^{2}}{d\hat{r}^{2}_{*}}+\tilde{V}_{\pm})\hat{G}^{\pm}=\omega^{2}\hat{G}^{\pm},
\end{equation}
where
\begin{equation}
\bar{V}_{\pm}=\frac{dW_{\pm}}{d\hat{r}_{*}}+W^{2}_{\pm},
\end{equation}
and
\begin{equation}
\tilde{V}_{\pm}=-\frac{dW_{\pm}}{d\hat{r}_{*}}+W^{2}_{\pm}.
\end{equation}

Here $k_\pm$ is related to the angular momentum quantum number $l$ as $k_{+}=l+1$ for $j=l+1/2$, and $k_{-}=-l$ for $j=l-1/2$. In the following sections we shall evaluate the QNMs in the case $j=l-1/2$, since the process is identical for $j=l+1/2$. Furthermore, since Dirac particles and antiparticles have the same QNMs in spherically symmetric regular spacetimes, the radial function $\hat{G}^-$ can represent all the relevant physics of Dirac field evolution in such spacetimes.

\section{Extreme charge in the spherically symmetric regular black holes}\label{sec:extremeCharge}

In order for the horizon to exist, $\beta$ should lie in the range $0 < \beta < \beta_e$, where $\beta_e$ is defined to be the extreme charge for which the inner horizon $r_p$ and outer horizon $r_+$ coincide. If $\beta$ exceeds $\beta_e$, the horizon would vanish, then the metric can not describe black hole spacetimes. Since the regular BHs have no singularity, it would not be restricted in Cosmic Censorship Hypothesis. Because the point of this paper is QNMs of BHs, We will not consider such cases in this research.

The concept of an extremal black hole is theoretical and none have thus far been observed in nature. However, many theories are based on their existence. Firstly extreme BHs have very simple physical properties, which can help us to study the theory of gravitation. Secondly their black hole entropy can be calculated by the Bekenstein-Hawking formula and string theory, which can solve the clash between string theory and Hawking's supporters. Last but not least, extreme black hole plays a very important role in the modern theory of the super-symmetric theory.

\begin{figure}
\centering \subfigure{
\begin{minipage}[t]{0.3\textwidth}
\includegraphics[width=1\textwidth]{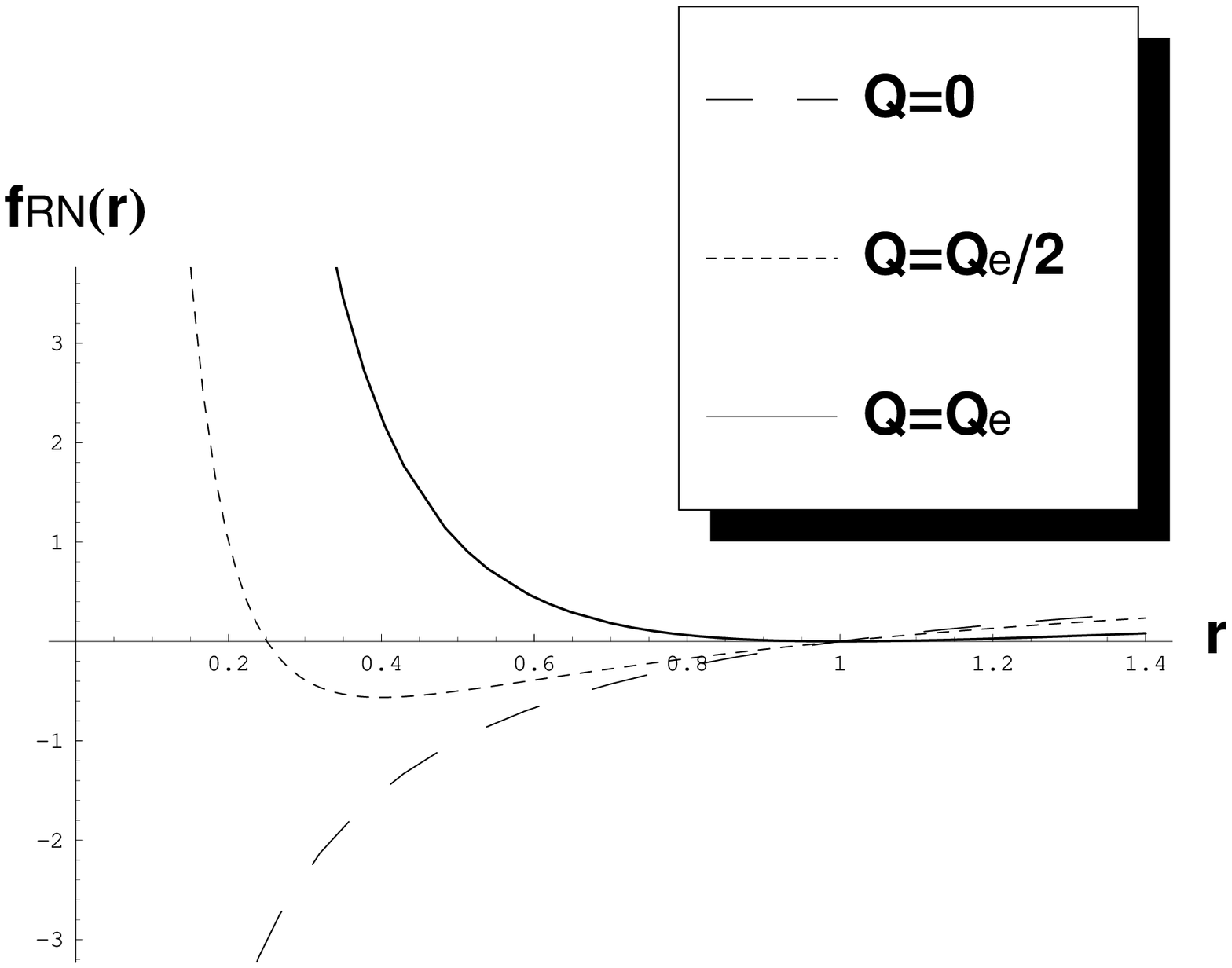} \\
\includegraphics[width=1\textwidth]{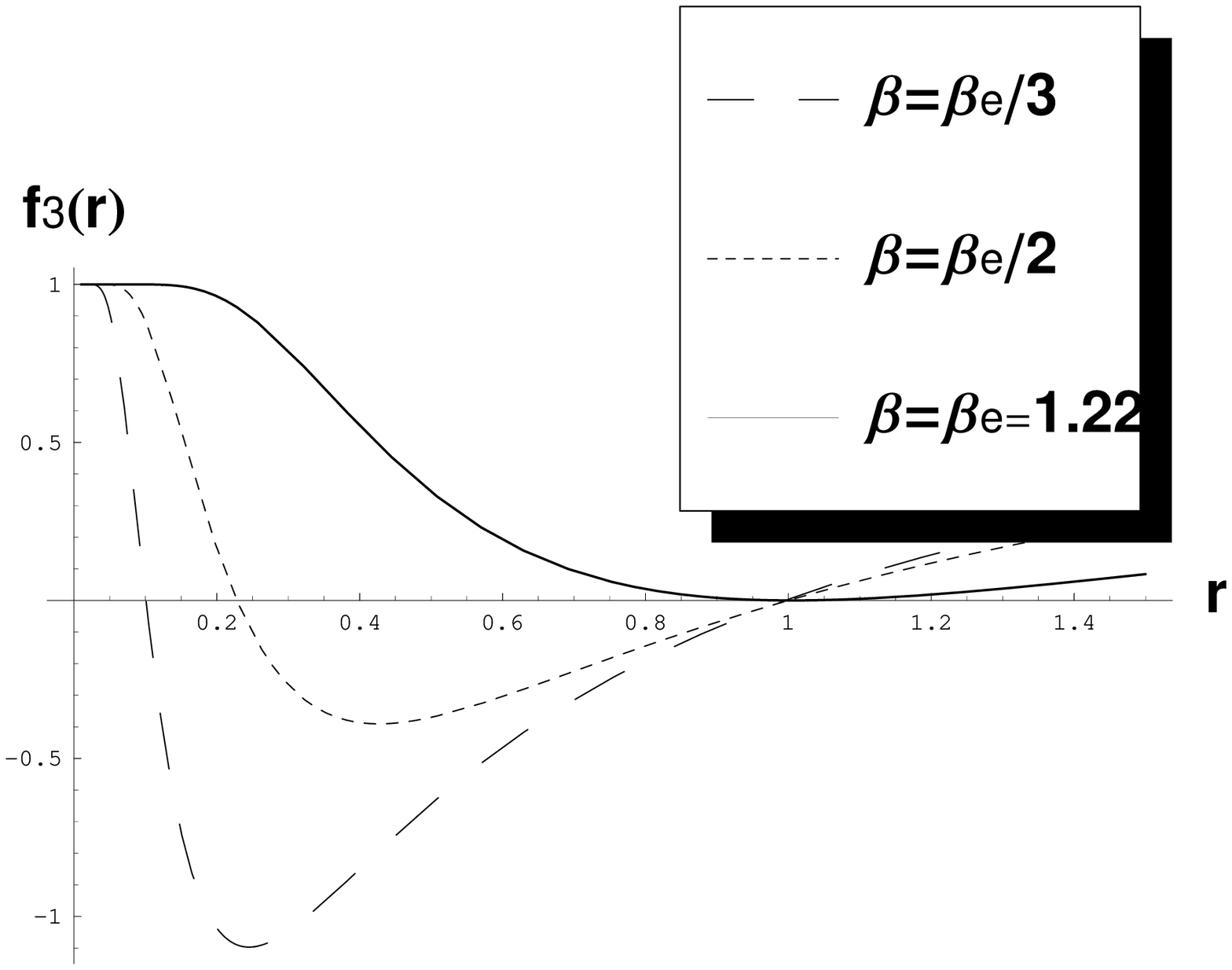}
\end{minipage}
} \subfigure{
\begin{minipage}[t]{0.3\textwidth}
\includegraphics[width=1\textwidth]{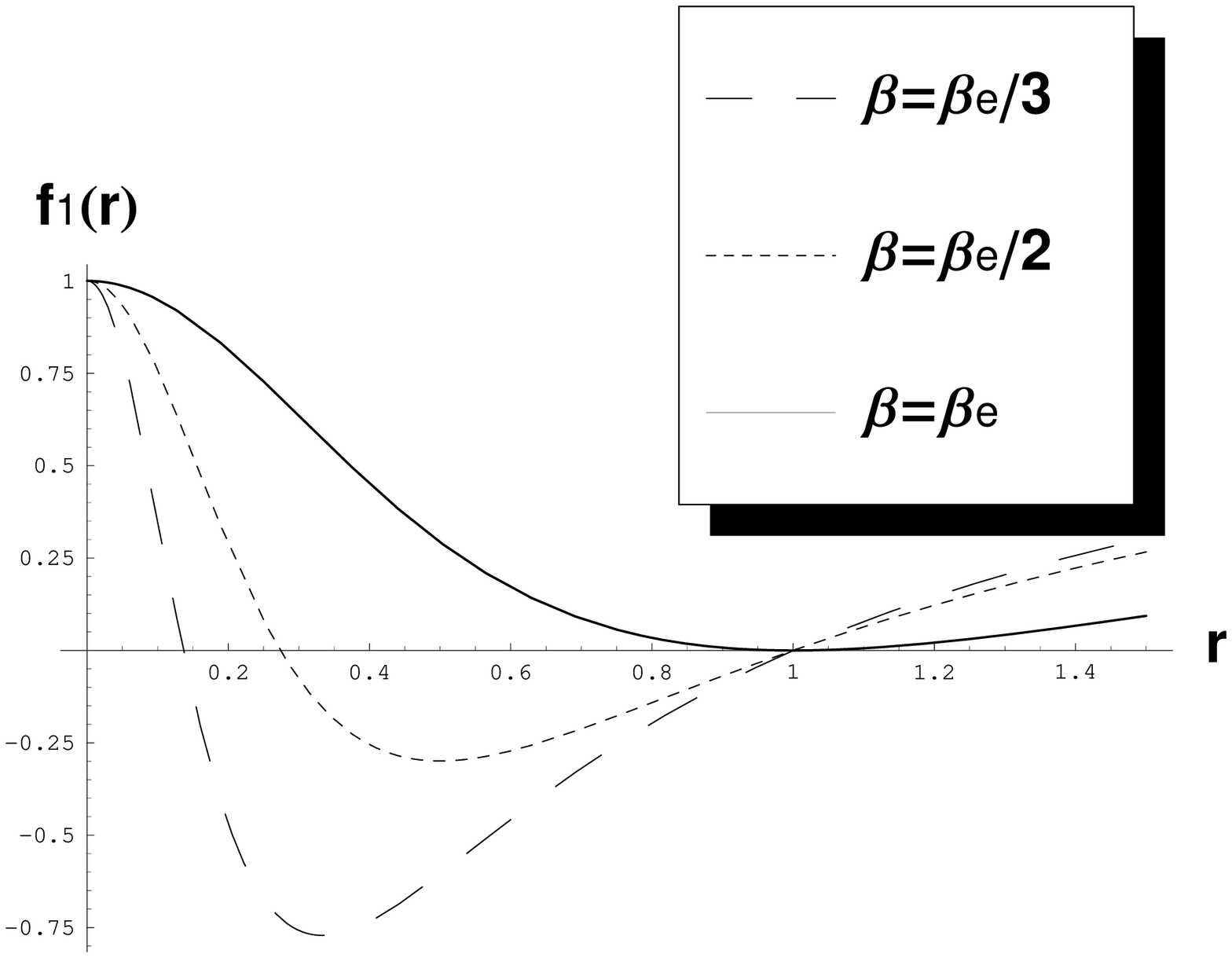} \\
\includegraphics[width=1\textwidth]{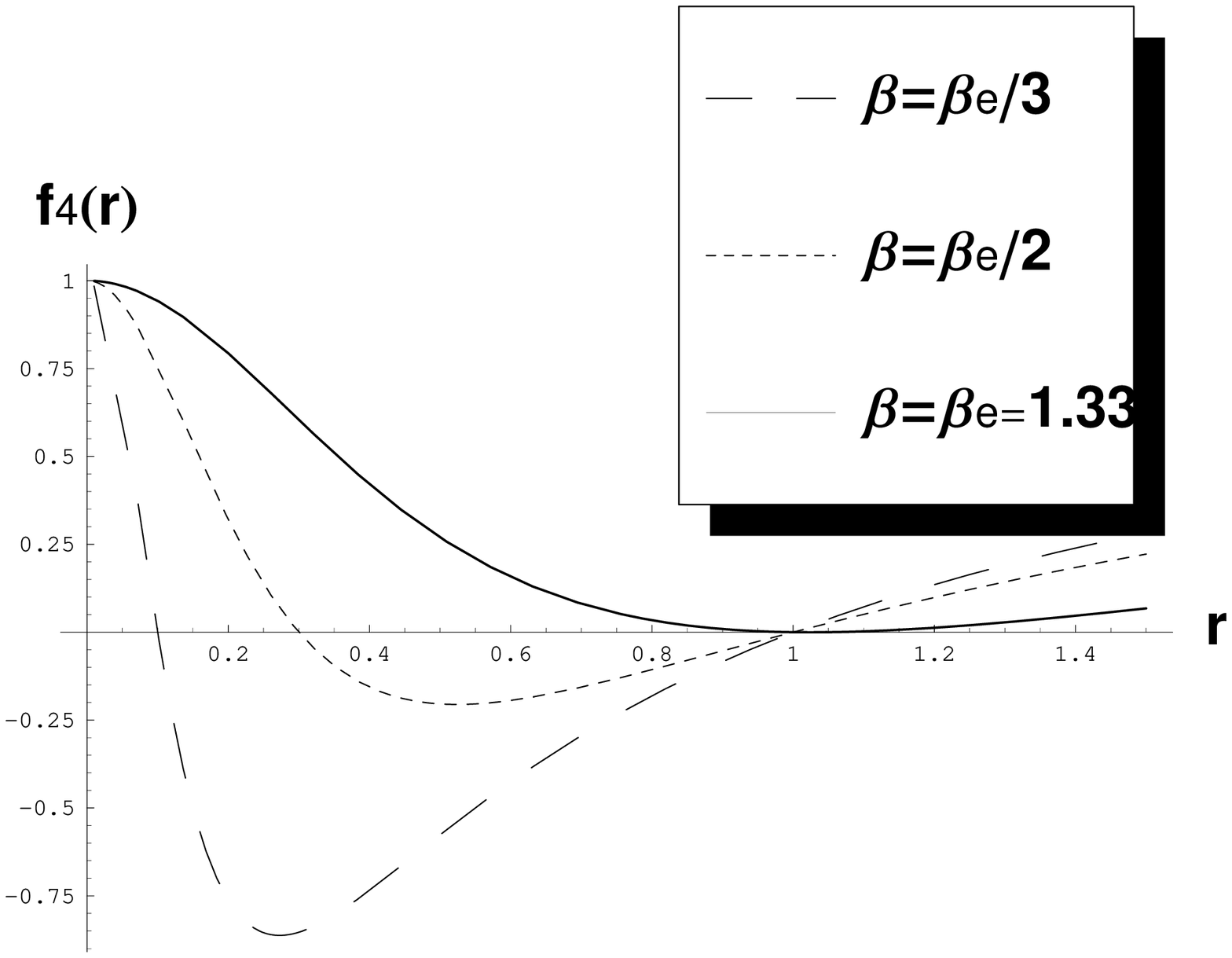}
\end{minipage}
} \subfigure{
\begin{minipage}[t]{0.3\textwidth}
\includegraphics[width=1\textwidth]{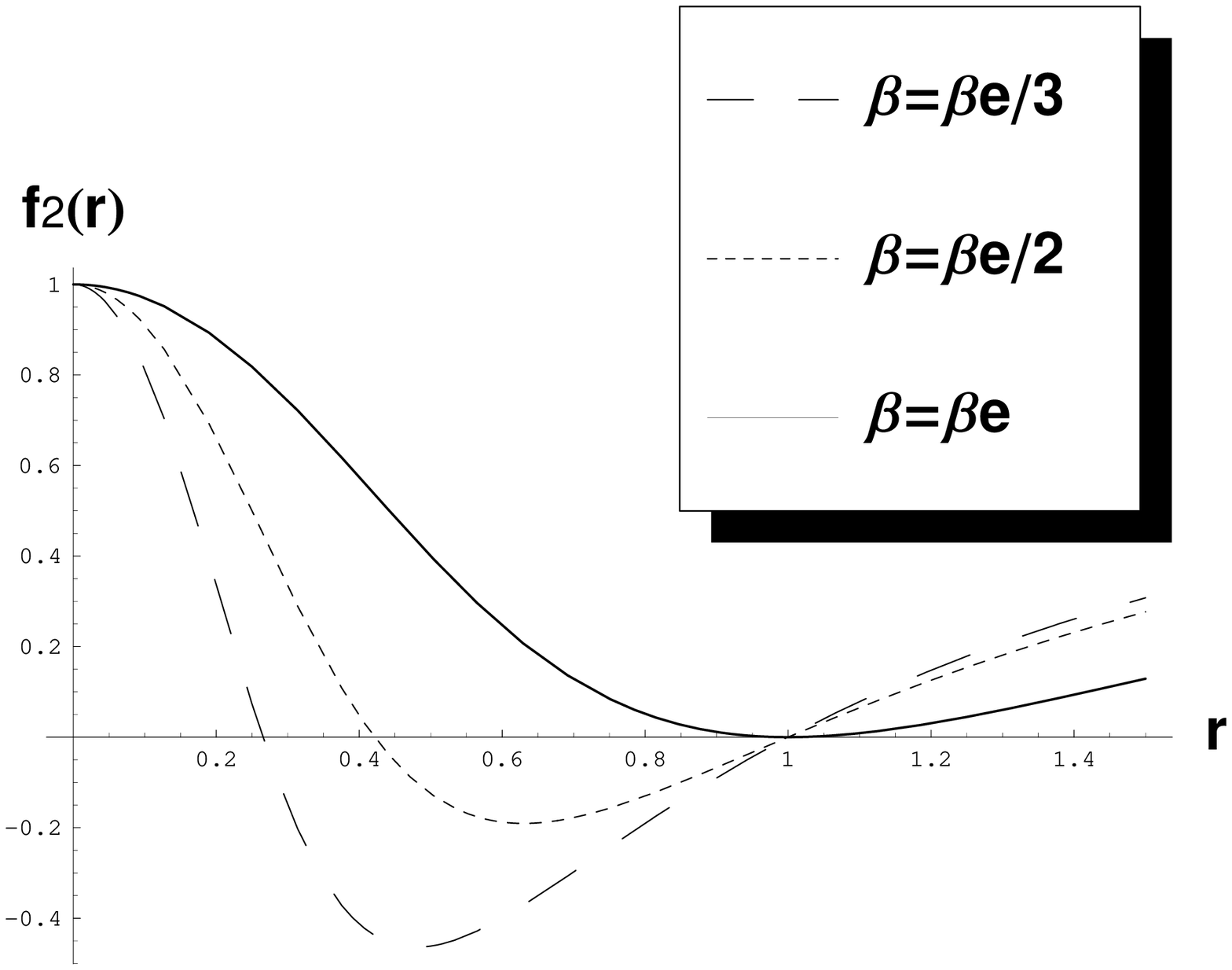} \\
\includegraphics[width=1\textwidth]{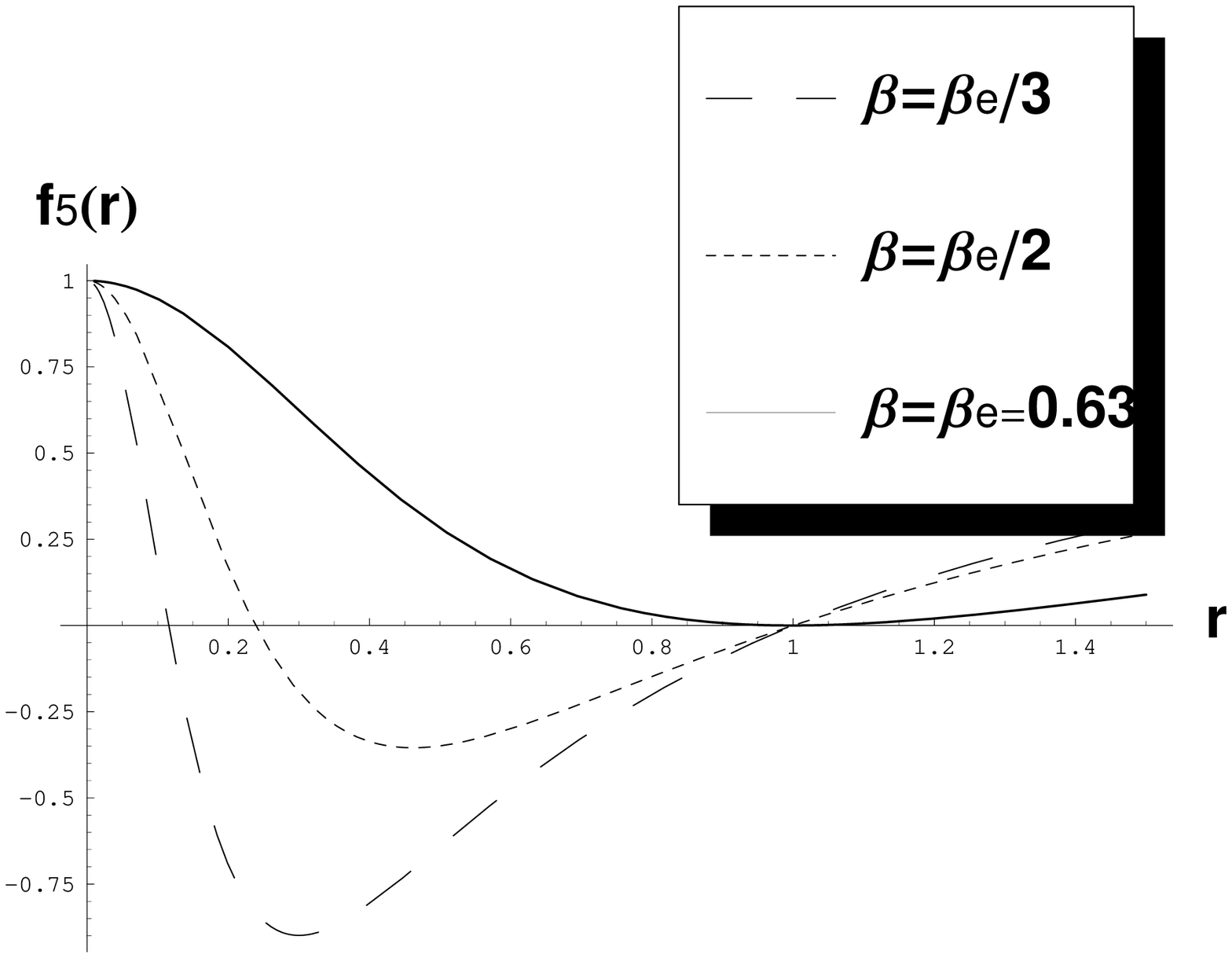}
\end{minipage}
}
\caption{The figure illustrates the behavior of the metric
functions $f_{RN}(r)(\text{top-left}),f_{1}(r)(\text{top-middle}),f_{2}(r)(\text{top-right})$, $f_{3}(r)(\text{bottom-left}),f_{4}(r)(\text{bottom-middle}),f_{5}(r)(\text{bottom-right})$ given $r_{+}=1$.}\label{fig:metrics}
\end{figure}

For comparison, we first consider the extreme condition for the Reissner-Nordstrom (RN) spherically symmetric spacetime. The RN metric can be written as
\begin{equation}
f(r)=\frac{(r-r_+)(r-r_p)}{r^{2}},
\end{equation}
\begin{equation}
M=\frac{Q^2+r_+^2}{2r_+};~~~~~~~~~Q_e^2=r_+^2;~~~~~~~~~~~r_p=Q^2/r_{+},
\end{equation}
where $Q_e$ is the extreme charge. From the first plot in Fig.~\ref{fig:metrics}, we see that $f_{\rm RN}(r)$ has a singularity at $r=0$. The equivalent Schwarzschild curve is shown for $Q=0$, and indicates one horizon where the curve crosses the horizontal axis. If $0 < Q < Q_e$, an inner horizon appears. At $Q=Q_e$, the inner and outer horizons coincide, so $r_p=r_+$ at $Q_e = r_+$.

We will similarly investigate the different $\beta_e$ for the regular spacetimes. Fortunately, for $f_1(r)$ and $f_2(r)$, the extreme charge can be analytically determined. For $f_1(r)$,
\bqn \label{Metric4}
M&=&\frac{(\beta^2+r_+^2)^{3/2}}{2r_+^2};~~~~~~~~~\beta_e^2=\frac{r_+^2}{2};\nb\\
r_p&=&\sqrt{\frac{\beta^6}{2r_+^4}+\frac{3\beta^4}{2r_+^2}+\frac{\beta^5\sqrt{\beta^{2}+4r_+^2}}{2r_+^4}+\frac{\beta^3\sqrt{\beta^{2}+4r_+^2}}{2r_+^2}}.\eqn
And rewrite $f_{2}(r)$ as
\begin{equation}
\label{Metric3}f(r)=\frac{(r-r_+)(r-r_p)(r-r_n)}{r^3+2\beta^2},
\end{equation}
where
\bqn \label{Metric4}
M&=&\frac{1}{2}\left(r_++\frac{2\beta^2}{r_+^2}\right);~~~~~~~~~\beta_e^2=\frac{r_+^3}{4};\nb\\
r_p&=&\frac{\beta^2+\sqrt{\beta^4+2\beta^2r_+^3}}{r_+^2};~~~~r_n=\frac{\beta^2-\sqrt{\beta^4+2\beta^2r_+^3}}{r_+^2},\eqn
then we find that $\beta_{e}=\sqrt{r_{+}^{2}/2}$ for $f_{1}(r)$ and $\beta_{e}=\sqrt{r_{+}^{3}/4}$ for $f_{2}(r)$.

For $f_{3}(r)$, $f_{4}(r)$ and $f_{5}(r)$, it is difficult to find an analytic relationship between $\beta_{e}$, $M$ and $r_{+}$, and $r_{p}$. Therefore, we use numerical methods to determine them from plots shown in figure~\ref{fig:metrics}. We see that the extreme charges should be $1.22 r_+$, $1.33 r_+$, and $0.63 r_+$ respectively. Note that the plots of $f_1(r)$ to $f_5(r)$ show similar behaviors. In particular, we note that $f(0)$ is finite in all cases---indicating regularity.

In the extreme condition, we can compute the QNM frequencies for the massless case using the WKB approximation. These results are shown in Table~\ref{tab:extremeQNMs}.

\begin{table}[!h]
\label{Table3} \tabcolsep 0pt \caption{QNMs frequencies on extreme condition evaluated by WKB Approach ($n=0$, $m=0$ and
$r_{+}=1$)}\label{tab:extremeQNMs}
\vspace*{-12pt}
\begin{center}
\def\temptablewidth{1\textwidth}
{\rule{\temptablewidth}{1pt}}
\begin{tabular*}{\temptablewidth}{@{\extracolsep{\fill}}cccc}
$f(r)$ &$|k|=1$ &$|k|=3$ &$|k|=3$\\
\hline
       $f_{RN}(r)$ &$0.2350-0.0889i$ &$0.4934-0.0883i$ &$0.7458-0.0882i$ \\
      $f_{1}(r)$ &$0.2226-0.0846i$ &$0.4738-0.0850i$ &$0.7173-0.0846i$ \\
      $f_{2}(r)$ &$0.2459-0.1090i$ &$0.5310-0.1082i$ &$0.8064-0.1079i$ \\
      $f_{3}(r) $ &$0.2136-0.0740i$ &$0.4478-0.0739i$  &$0.6765-0.0736i$ \\
      $f_{4}(r) $ &$0.2029-0.0679i$ &$0.4249-0.0682i$  &$0.6418-0.0680i$ \\
      $f_{5}(r) $ &$0.2191-0.0805i$ &$0.4641-0.0803i$  &$0.7021-0.0801i$ \\

       \end{tabular*}
       {\rule{\temptablewidth}{1pt}}
       \end{center}
\end{table}

\section{QNMs for the massless Dirac field}\label{sec:masslessDirac}

\begin{figure}
\centerline{\includegraphics[height=5cm]{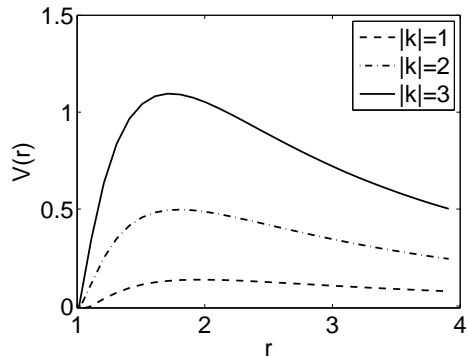}}
 \caption{Variation of the effective potential $V(r)$
with respect to the polar coordinate $r$ for the five regular
space-times, In the plot, the solid, dot-line, dashing lines
correspond to $|k|=1,2,3$ respectively. We set $\beta=\beta_{e}/2,r_{+}=1$.}\label{fig:effectivePot}
\end{figure}

In order to compute the QNM frequencies, we first need to determine the properties of the effective potential $V(r,|k|)$. In the massless case, the Schr\"{o}dinger-like equation in the stationary state for $\hat{G}^{\pm}$ and $\hat{F}^{\pm}$
 can be simplified. As mentioned earlier, we use $\hat{G}^-$ as the example, thus

\begin{equation}
\label{waveeq1}
(-\frac{d^{2}}{d\hat{r}^{2}_{*}}+V)\hat{G}^{-}=\omega^{2}\hat{G}^{-},
\end{equation}
\label{eq.1}
 where
\begin{equation}
\label{waveeq2}
V=\tilde{V}_{-}|_{m=0}=-\frac{dW_{-}}{d\hat{r}_{*}}+W^{2}_{-},
\end{equation}
and
\begin{equation}
\label{waveeq3}
W_{-}=\sqrt{f(r)}\frac{|k_{-}|}{r}.
\end{equation}
In order to simplify the notation, we drop the `$-$' subscripts and superscripts for the remainder of this section. We also set $r_+ = 1$. Fig.~\ref{fig:effectivePot} illustrates the potential's behavior and shows that the peak of the potential barrier increases as $|k|$ increases, while the location of the peak moves closer to $r=0$.

\begin{figure}
\centering \subfigure{
\begin{minipage}[t]{0.3\textwidth}
\includegraphics[width=1\textwidth]{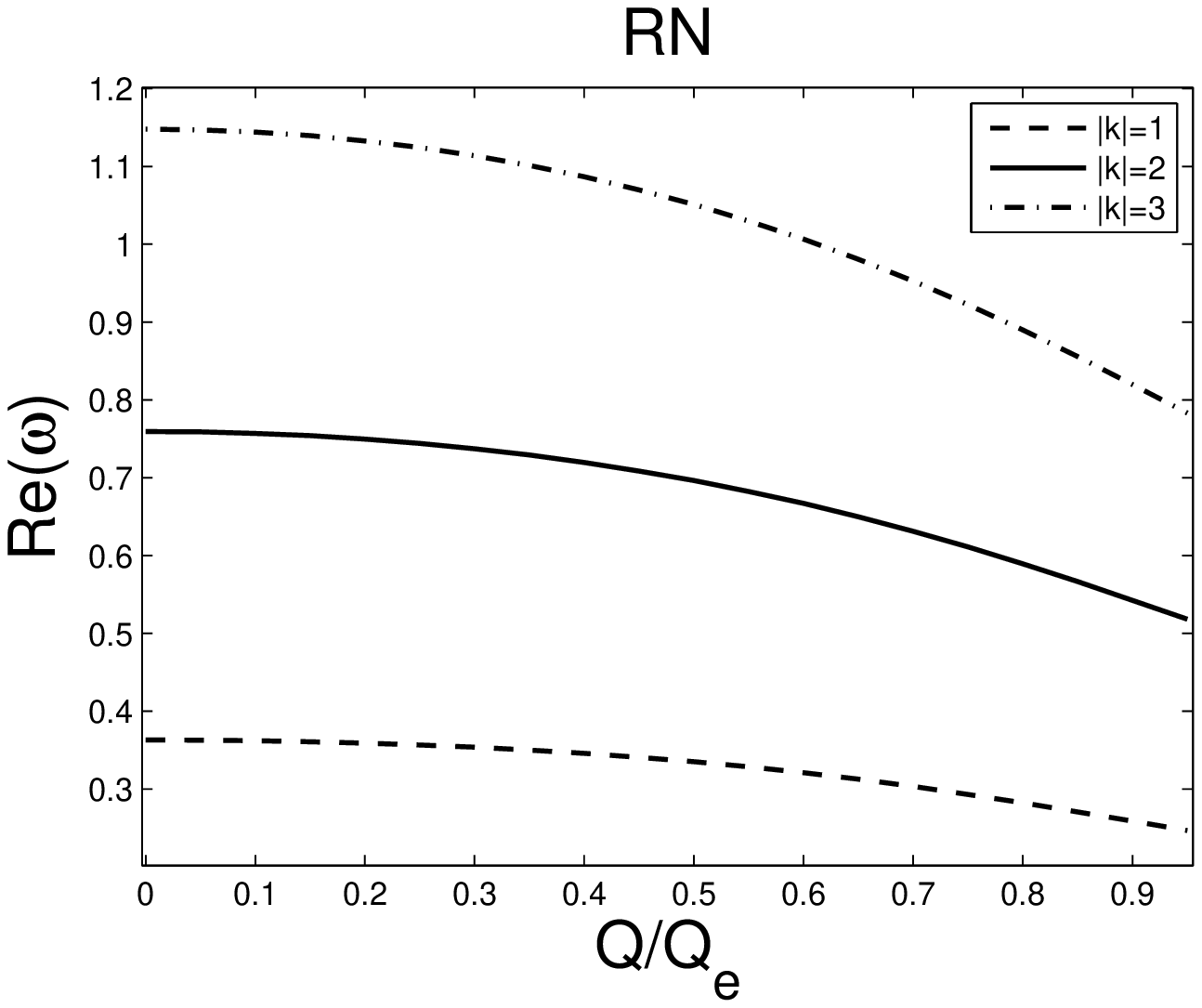} \\
\includegraphics[width=1\textwidth]{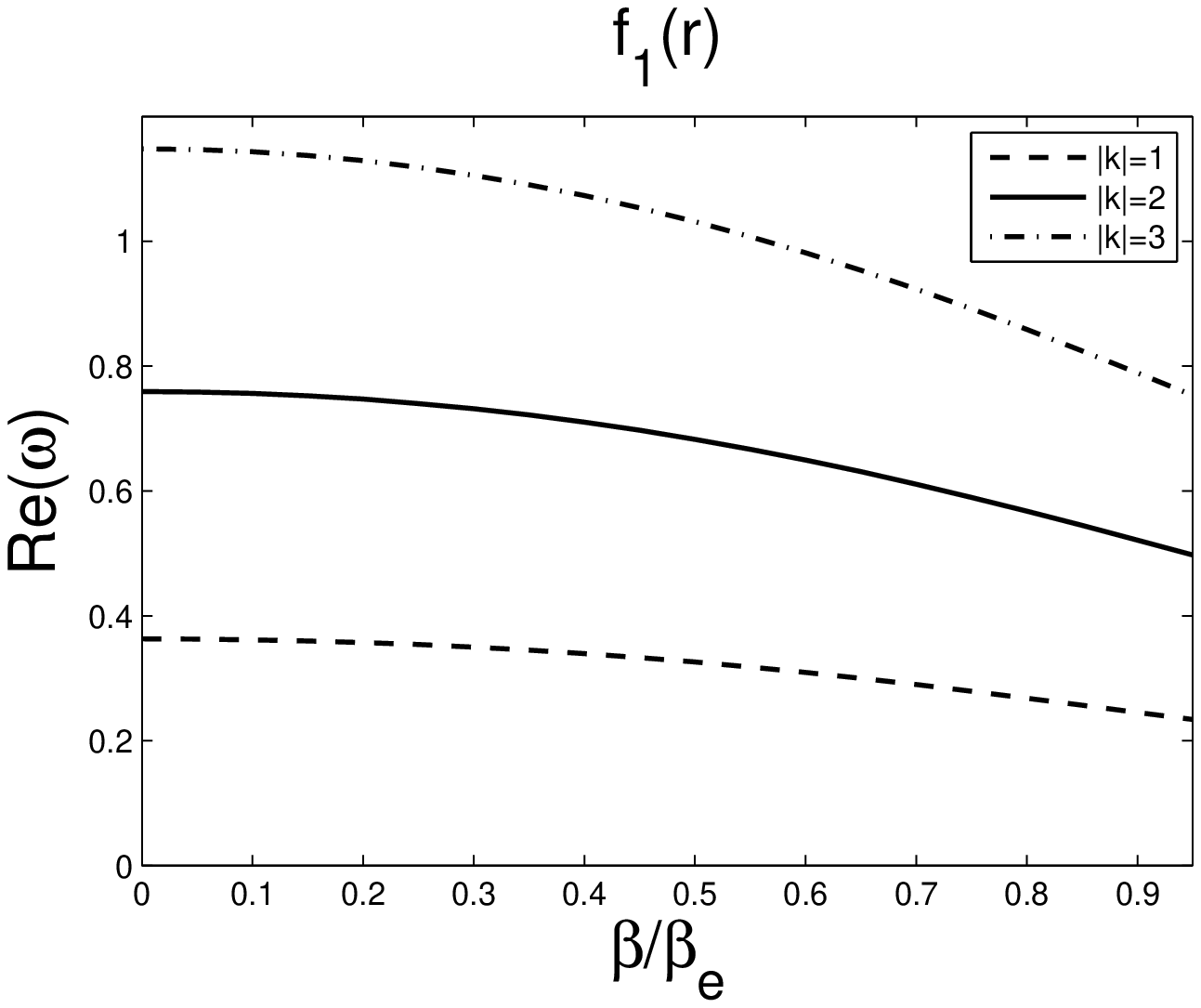} \\
\includegraphics[width=1\textwidth]{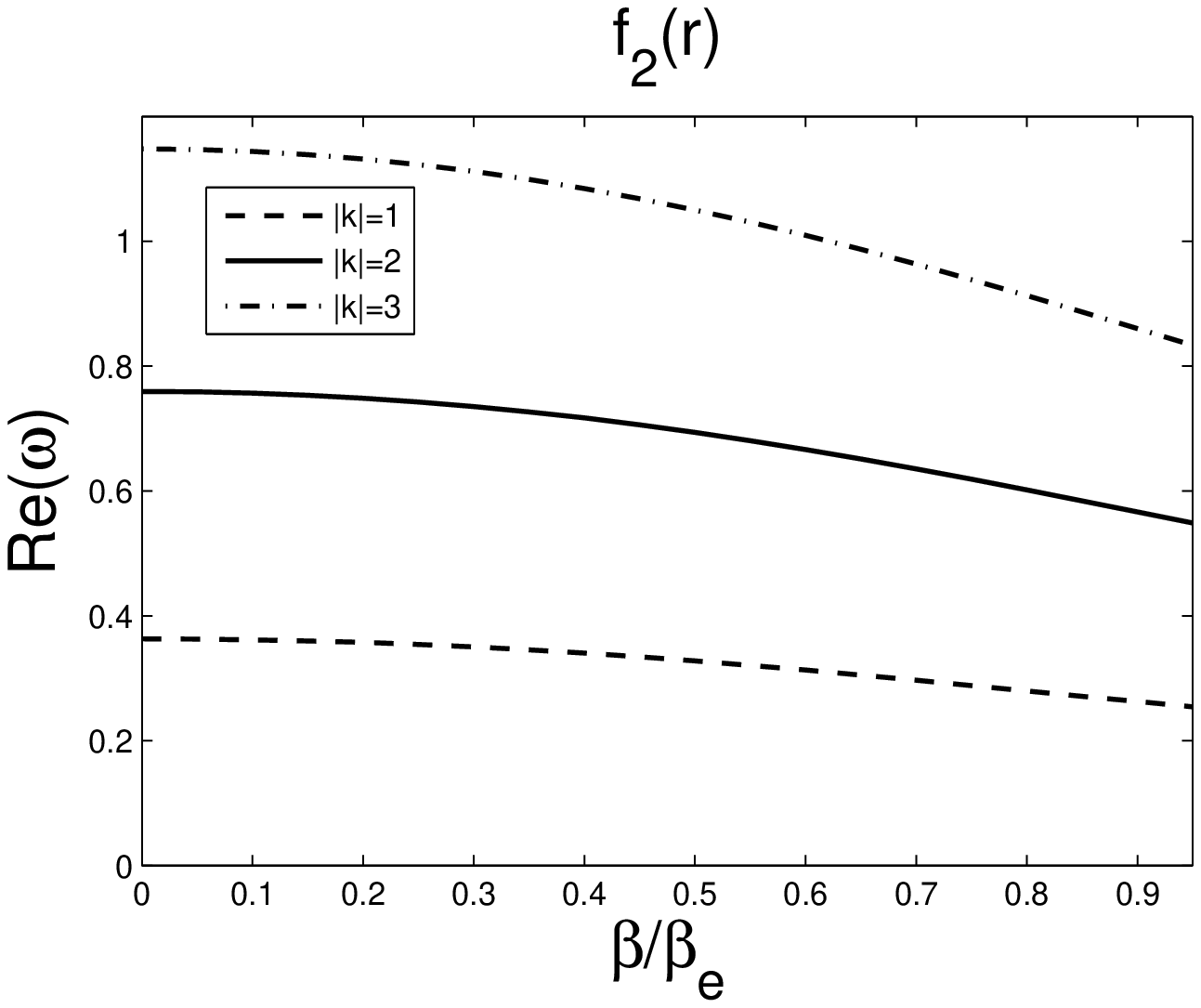} \\
\includegraphics[width=1\textwidth]{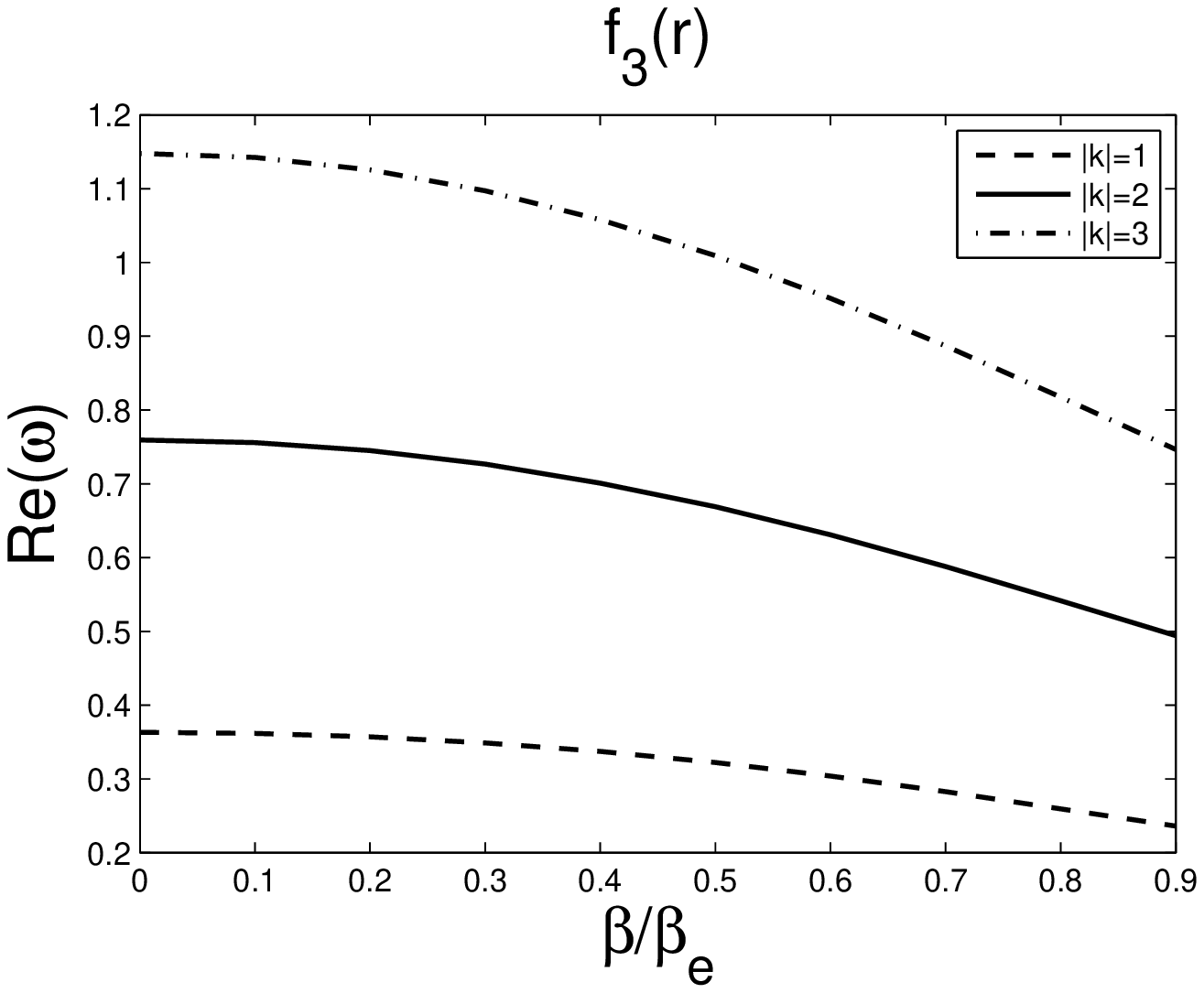} \\
\includegraphics[width=1\textwidth]{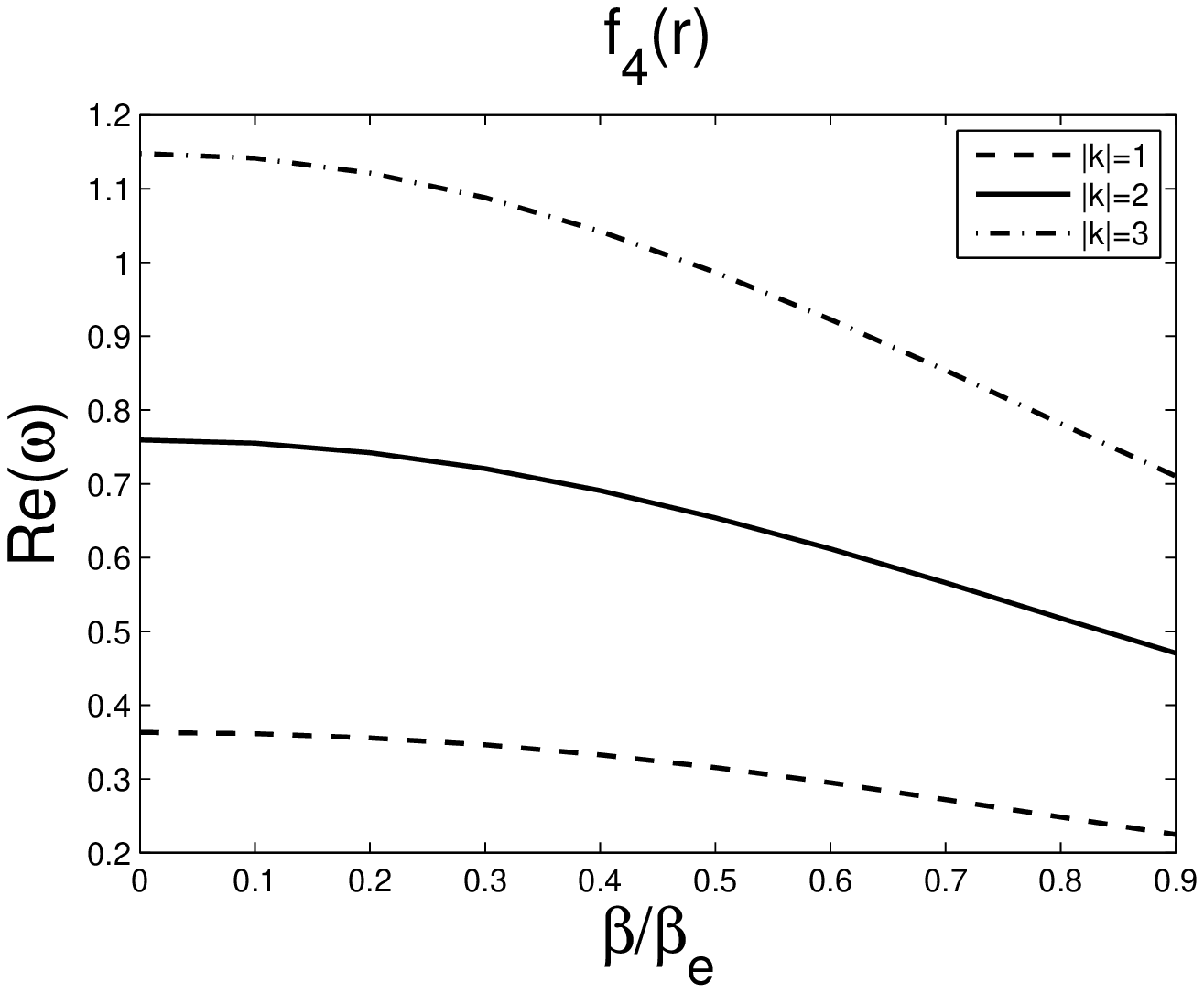} \\
\includegraphics[width=1\textwidth]{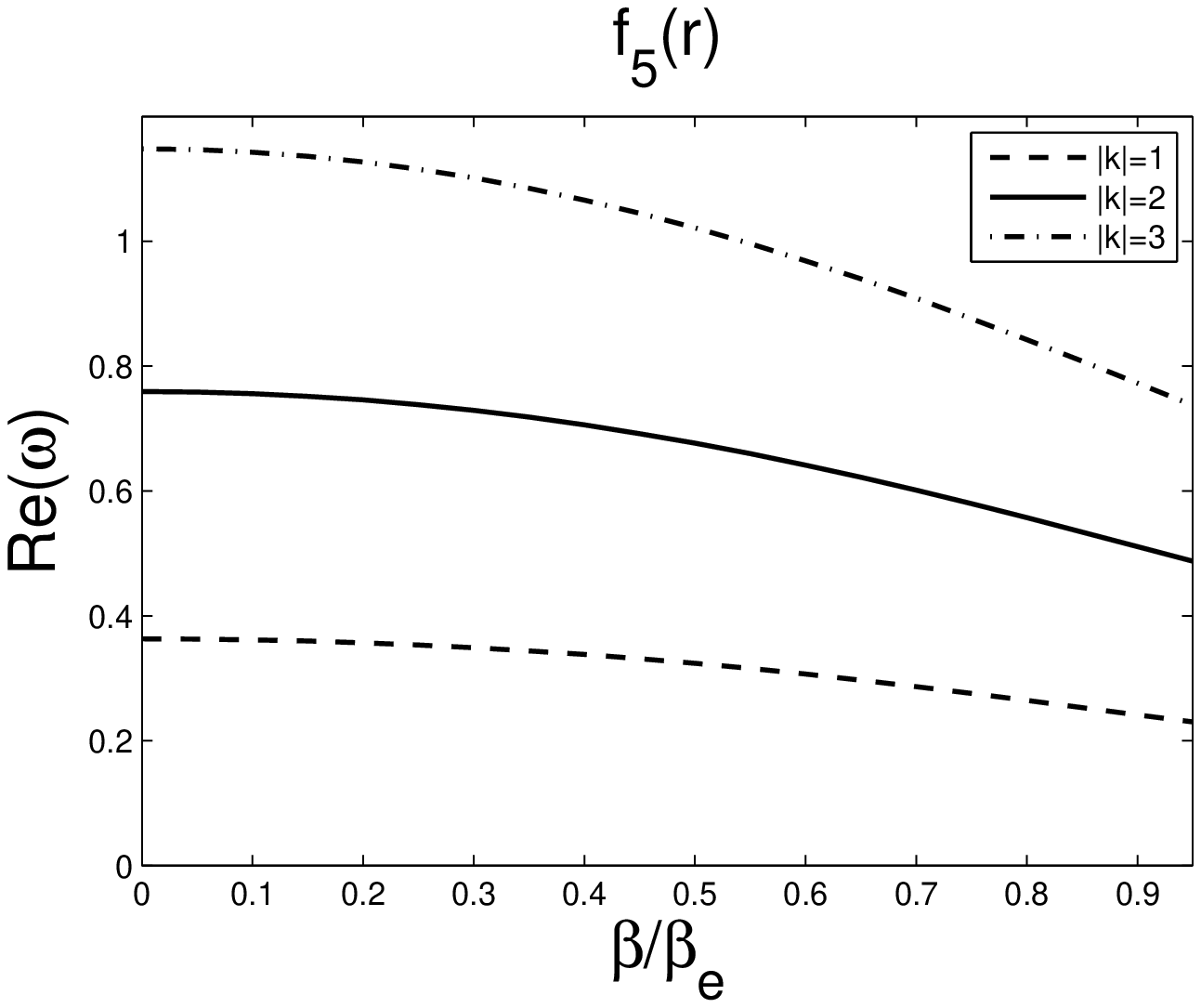}
\end{minipage}
} \subfigure{
\begin{minipage}[t]{0.3\textwidth}
\includegraphics[width=1\textwidth]{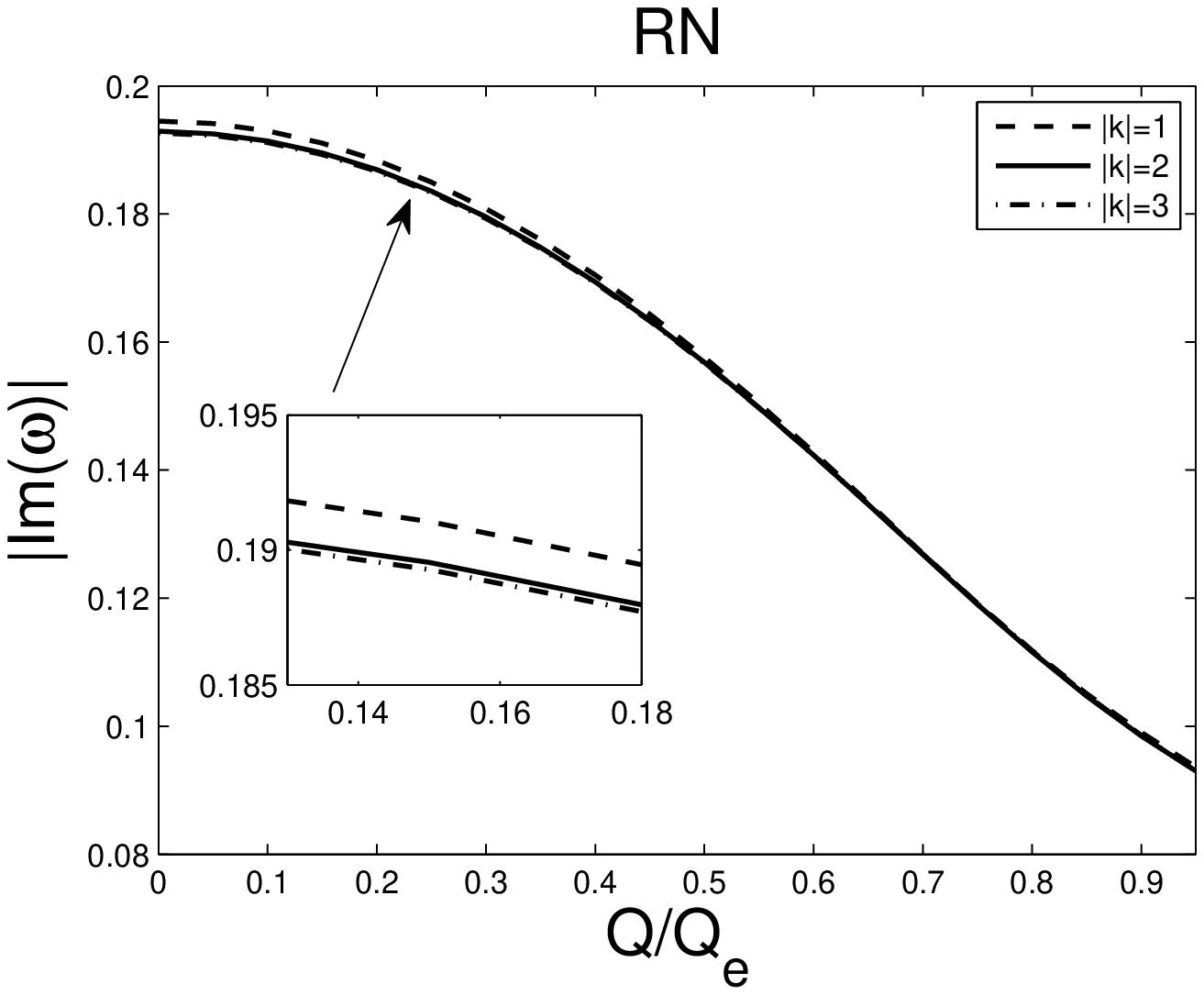} \\
\includegraphics[width=1\textwidth]{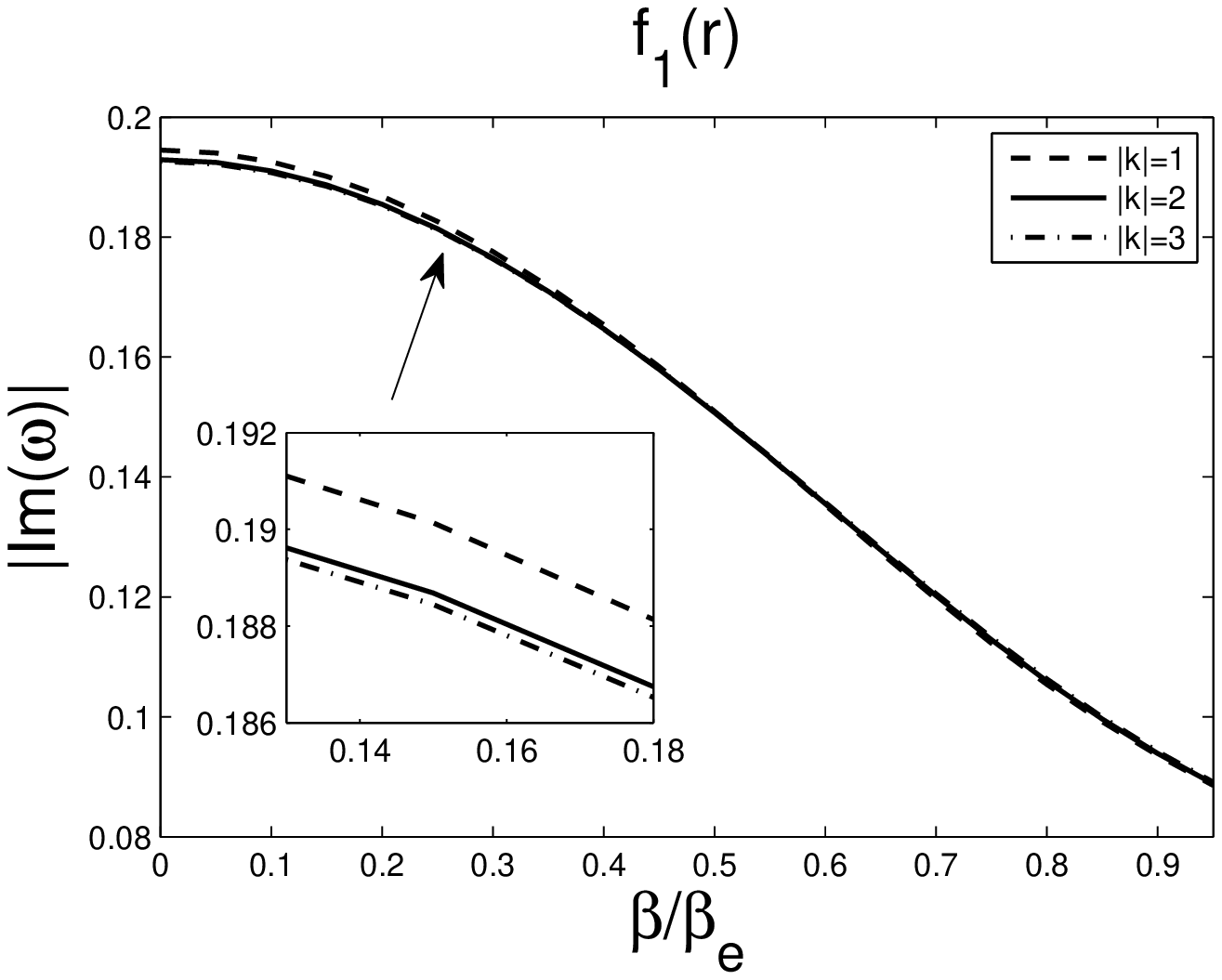} \\
\includegraphics[width=1\textwidth]{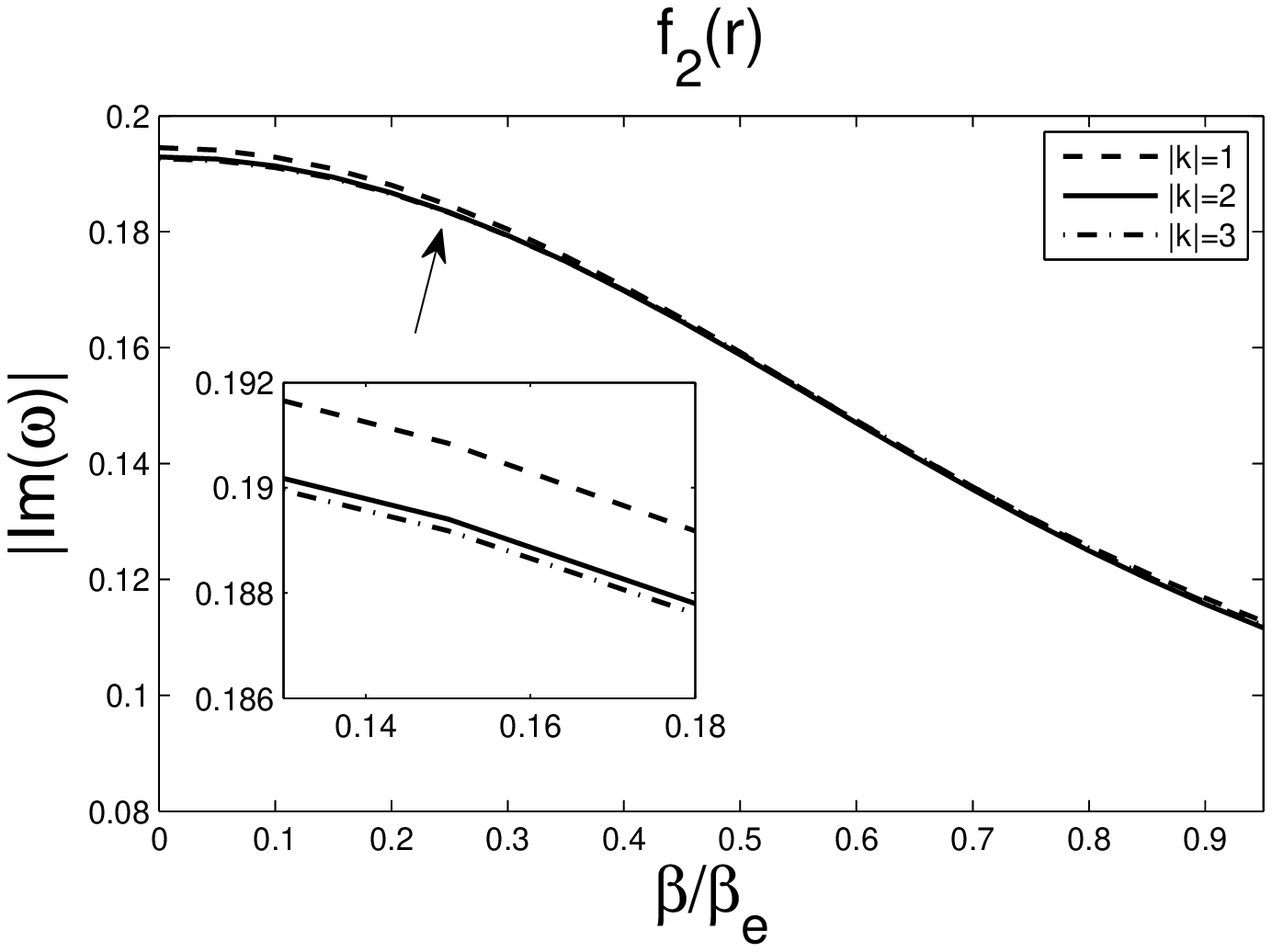} \\
\includegraphics[width=1\textwidth]{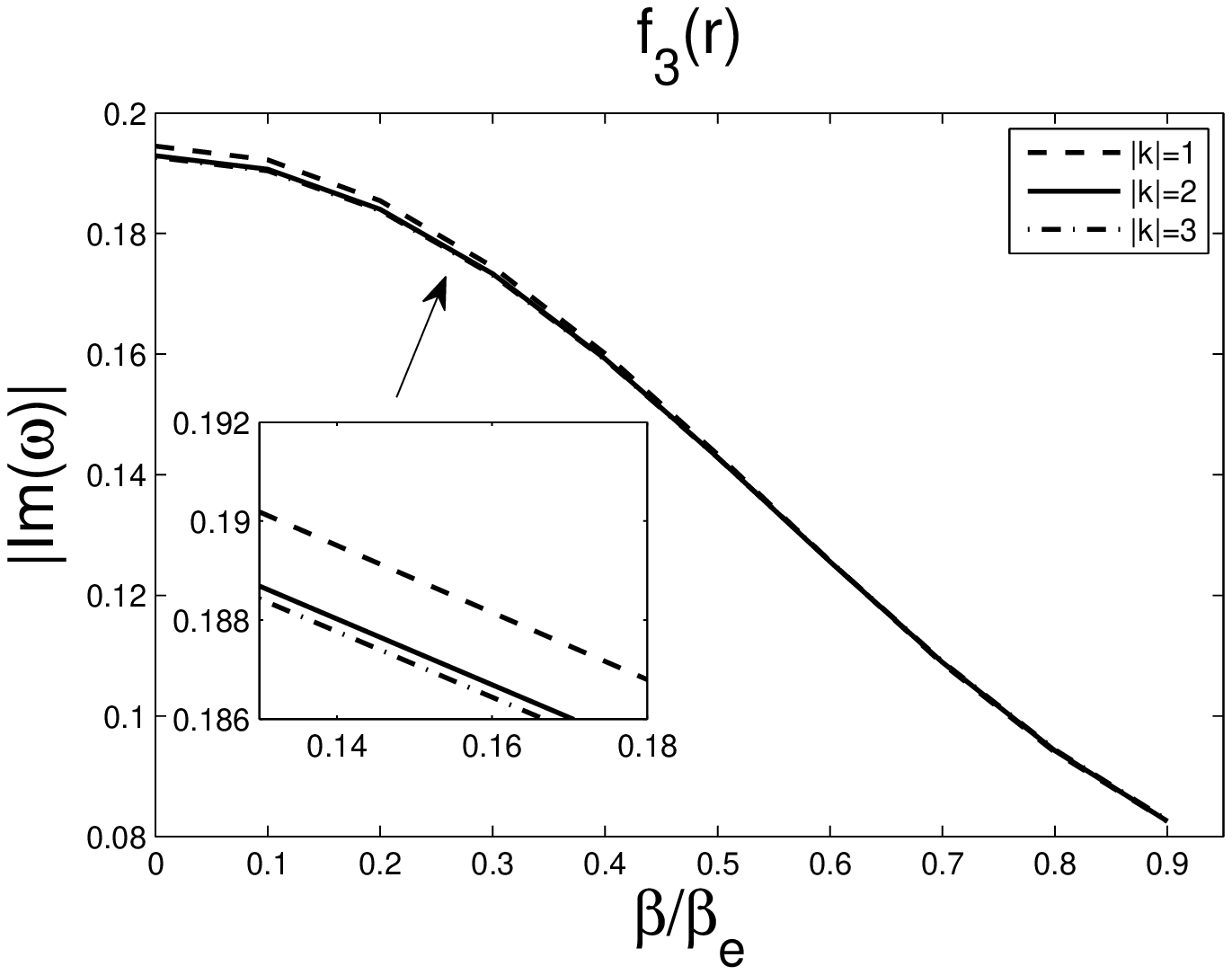} \\
\includegraphics[width=1\textwidth]{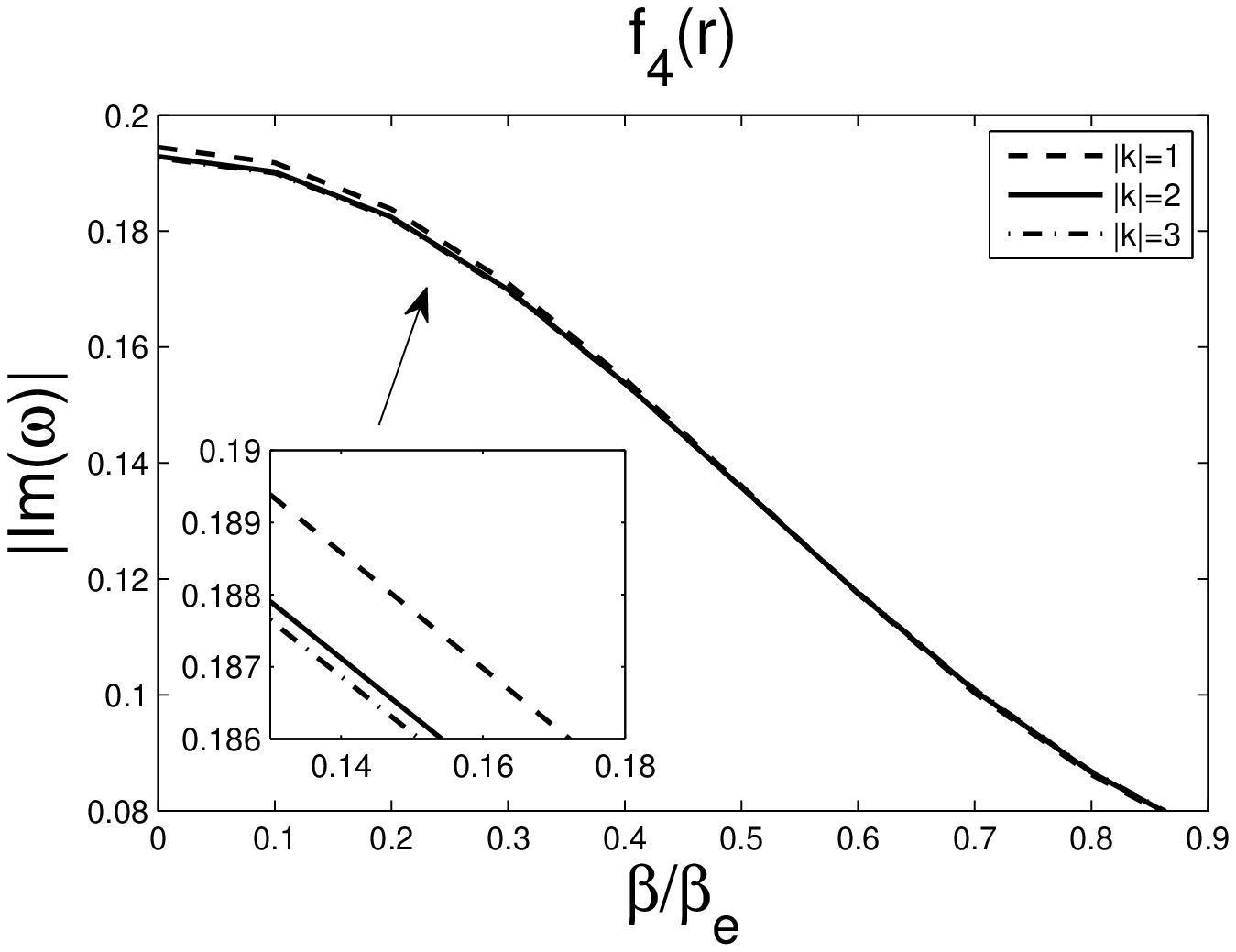} \\
\includegraphics[width=1\textwidth]{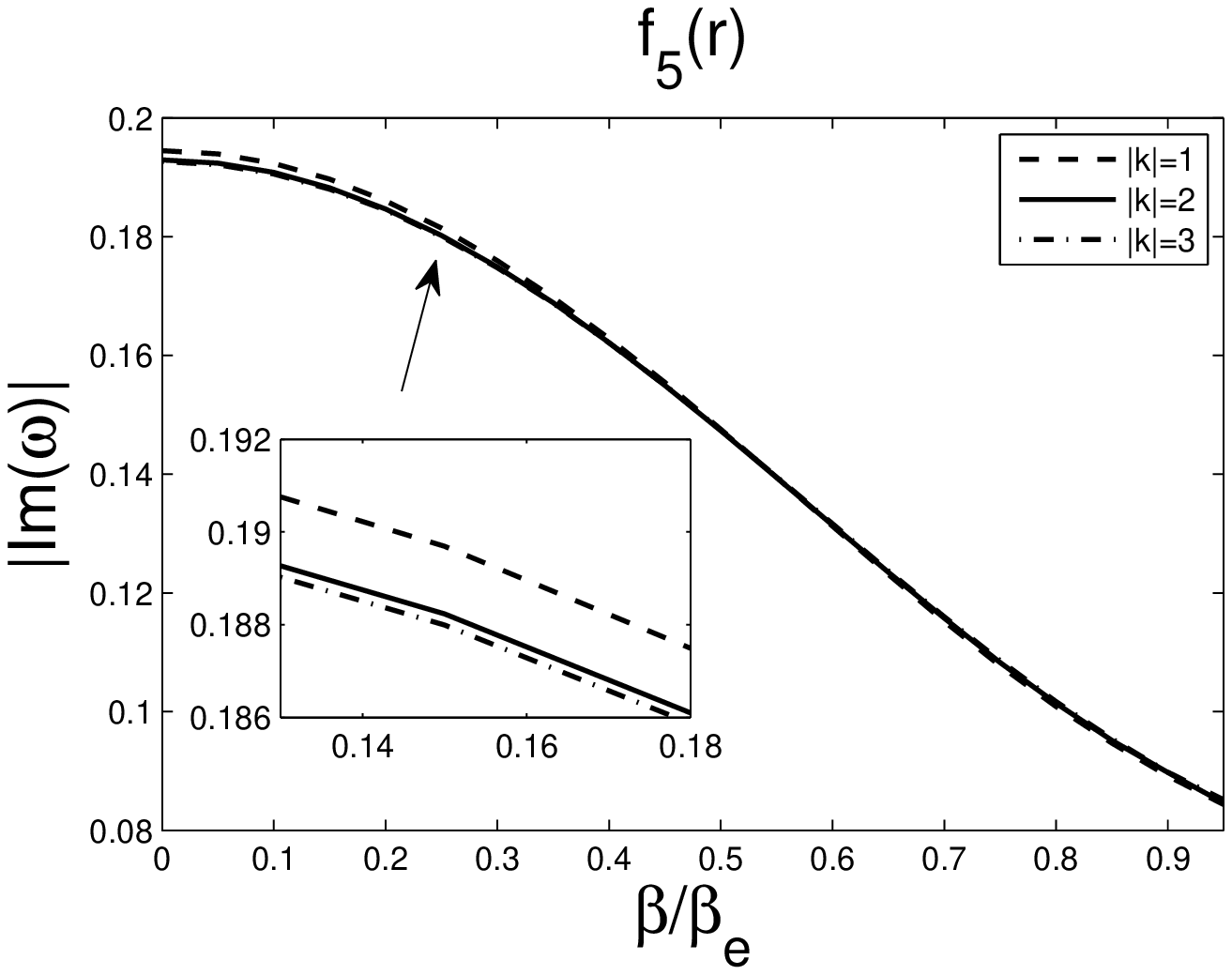}
\end{minipage}
}
\caption{Massless Dirac QNMs frequencies with various
$\beta$.}\label{fig:masslessFreqs}
\end{figure}

We use the 3$^{rd}$-order WKB approximation to calculate the QNM frequencies for RN BH and regular BHs. The key equations for evaluating the complex frequencies were given in~\cite{WKB1}, and depend on the potential and its derivatives at the peak. The QNM frequencies of each $|k|$ are plotted over the range of charge in Fig.~\ref{fig:masslessFreqs}, with $r_+=1$ and $n=0$. There are common behaviors of the QNM frequency for all $f(r)$ studied. From the plots in Fig.~\ref{fig:masslessFreqs}, we see that ${\rm Re}(\omega)$ increases with increasing $|k|$ while charge is held constant, and decreases with increasing charge while $|k|$ is held constant. Furthermore, $|{\rm Im}(\omega)|$ decreases significantly with increasing charge, while large $|k|$ results in slower decay. This indicates that the size of the magnetic monopole charge influences the behavior of QNM decay. Both $|k|$ and charge are related to the description of the evolution of the Dirac field in these regular spacetimes. In particular, the regular BHs with $\beta=0$ would return to Schwarzschild case, which occur more prominent damping and quicker oscillation rate. Those properties provide a way to distinguish between a Schwarzschild BH, and any one of the five regular BH models.

Next, using the expansion method, the QNM frequencies can be found by searching for a single solution that is valid everywhere outside the horizon. Since the QNM frequencies are related to the orbital frequency and the Lyapunov exponents for geodesics at the unstable orbit, the key step is to find the radius of the unstable circular orbit. Once a correct radius has be found, the QNM frequencies may be expressed as expansions in inverse powers of $L=l+1/2$.

In our case, the Dirac perturbation is governed by the wave equation (see Eq.(\ref{waveeq1})-Eq.(\ref{waveeq3})),
\begin{equation}
\left[\frac{d^{2}}{d\hat{r}^{2}_{*}}+\omega^{2}-V(r,|k|)\right]\hat{G}(r)=0.
\end{equation}
The solution satisfies the following boundary conditions:
\begin{enumerate}
\item Pure ingoing waves at the
event horizon $\hat{G}\sim e^{-i\omega \hat{r_{*}}}$,
$\hat{r}_{*}\rightarrow -\infty$.
\item Pure outing waves at the
spatial infinity $\hat{G}\sim e^{i\omega
\hat{r_{*}}}$,$\hat{r}_{*}\rightarrow \infty$.
\end{enumerate}
Following Dolan and Ottewill, the function of radial
equation $\hat{G}(r)$ can be redefined as
\begin{equation}
\hat{G}(r)=\upsilon(r)exp{(\int^{\hat{r}_{*}}\alpha(r)d\hat{r}_{*})},
\end{equation}
where $\alpha(r)=i\omega b_{c}k_{c}(r)$ and
\begin{equation}
k_{c}(r)=(r-r_{c})\sqrt{\frac{\hat{k}^{2}(r,b_{c})}{(r-r_{c})^{2}}},\\
\hat{k}^{2}(r,b)=\frac{1}{b^{2}}-\frac{f(r)}{r^{2}},
\end{equation}
with the condition
\begin{equation}
\hat{k}^{2}(r_{c},b_{c})=\frac{\partial \hat{k}^{2}(r,b_{c})}{\partial
r}|_{r=r_{c}}=0.
\end{equation}
For the massless Dirac field, $\frac{d}{d\hat{r}_{*}}=\frac{d}{dr_{*}}$,
yielding the Dirac wave equation
\begin{equation}
\label{eikonal4}
\frac{d^{2}\upsilon(r)}{dr_{*}}+2\alpha(r)f(r)f'(r)+[\omega^{2}+\alpha^{2}(r)-V(r)+f(r)\alpha'(r)]\upsilon(r)=0.
\end{equation}
For the fundamental mode $n=0$,  $\omega$ and $\upsilon(r)$ can
be expanded as
 \bqn
\label{eikonal5}
\omega=\sum\limits_{i=-1}^{\infty}\left(\frac{a_i}{L^i}\right),~~~~~~
\ln v(r)=\sum\limits_{i=0}^{\infty}\left[L^{-i}S_i(r)\right]. \eqn

Finally, we expand eq.(\ref{eikonal4}) in powers of $L^{-1}$. For our purpose, the $a_i$ need to be determined so that the QNM frequencies can be evaluated. A more detailed discussion of this approach can be found in~\cite{RegularQuasinormal1} and~\cite{eikonal2}. He we focus on the massless Dirac field perturbation in regular BHs. The QNM frequencies determined by both the WKB and expansion methods are given in Table~\ref{tab:expansionWKB}

\begin{table}[!h]
 \tabcolsep 0pt \caption{QNMs frequencies evaluated by WKB Approach and expansion method ($n=0$, $m=0$, $\beta=\beta_{e}/2$ and
$r_{+}=1$)}\label{tab:expansionWKB}
 \vspace*{-12pt}
\begin{center}
\def\temptablewidth{1\textwidth}
{\rule{\temptablewidth}{1pt}}
\begin{tabular*}{\temptablewidth}{@{\extracolsep{\fill}}cccc}
$f(r)$ &$|k|$ &$\text{WKB}~\text{Approach}~(\text{3}~\text{order})$ &$\text{expansion}~\text{method} ~(\text{ignore}~L^{-3}~\text{term})$ \\
\hline
      $ $ &$1$ &$0.326094-0.150958i$  &$0.332569-0.150893i$ \\
      $f_{1}(r)$ &$2$ &$0.682782-0.150855i$ &$0.684157-0.150839i$ \\
      $ $ &$3$ &$1.03151-0.150752i$  &$1.032-0.150830i$  \\
      \hline
      \hline
      $ $ &$1$ &$0.327835-0.159209i$  &$0.33574-0.159109i$ \\
      $f_{2}(r)$ &$2$ &$0.693009-0.159028i$ &$0.695674-0.158913i$ \\
      $ $ &$3$ &$1.05016-0.158783i$  &$1.05085-0.15820i$  \\
      \hline
       \hline
      $$ &$1$ &$0.322384-0.143348i$ &$0.327053-0.142963i$ \\
      $f_{3}(r) $ &$2$ &$0.668827-0.142878i$  &$0.669774-0.142903i$ \\
      $ $ &$3$ &$1.0091-0.14285i$  &$1.00941-0.142886i$  \\
      \hline
       \hline
      $$ &$1$ &$0.315488-0.136083i$ &$0.320246-0.13595i$ \\
      $f_{4}(r) $ &$2$ &$0.654051-0.135784i$  &$0.654979-0.135919i$ \\
      $ $ &$3$ &$0.986598-0.135781i$  &$0.986861-0.135910i$  \\
      \hline
       \hline
      $$ &$1$ &$0.323952-0.147578i$ &$0.329926-0.147437i$ \\
      $f_{5}(r) $ &$2$ &$0.67649-0.147409i$  &$0.677742-0.147401i$ \\
      $ $ &$3$ &$1.0216-0.147337i$  &$1.02204-0.147272i$  \\
       \end{tabular*}
       {\rule{\temptablewidth}{1pt}}
       \end{center}
\end{table}
We find that the precision of both of these approaches is adequate for regular spacetimes. Since the expansion method is based on the eikonal limit, better agreement between approaches occurs as $|k|$ (i.e. $l$) increases. In fact, the expansion of the nodes with $n>0$ discussed in~\cite{eikonal2} are less accurate, so we don't consider these cases further in this paper.

\section{QNMs for the massive Dirac field}\label{sec:massiveDirac}

\begin{figure}
\centerline{\includegraphics[height=5cm]{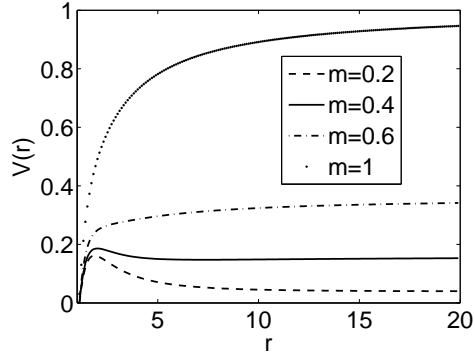}}

\caption{The effective potential for the massive Dirac
field. Here $|k|=1,\beta=\beta_{e}/4,r_{+}=1$.}\label{fig:massivePotential}
\end{figure}

In the massive case, the effective potential function $V(r)$ depends on $m$ as well as $\omega$. This makes the calculation much more complicated. We show in figure~\ref{fig:massivePotential} the dependence of $V(r)$ on $m$ for $\omega=1$. It is found that in all the regular spacetimes considered, $V(r)$ exhibits the following behavior
\begin{equation}
V(r\rightarrow\infty)=m^{2}.
\end{equation}
For small masses, $V(r)$ still has the form of a barrier potential, but with increasing mass, the peak of the potential increases slowly enough that eventually the height of the peak is lower than the asymptotic value of $m^2$. Further increases of $m$ turns the potential barrier into a potential step. However, $\omega$ is actually unknown at the start of the calculation and so it is probably not 1 and must be determined self-consistently. Hence, we should consider figure~\ref{fig:massivePotential} only to be indicative of the general behavior of $V$.

Using the WKB method to calculate the QNM frequencies, we find that the real part of $\omega$ increases with increasing $m$ while $|k|$ is held constant, while the imaginary part decreases with increasing $m$. This indicates that the QNMs for more massive field particles decay slower. On the other hand, given the mass of field $m$ and overtone $n$, with adjacent multiple $|k|$ increase the oscillation becomes more tenser and damping slower. Meanwhile the first overtone mode with fixed $m$ and $|k|$ tends to decay quicker than the fundamental one. For any $m$, the fundamental mode ($n=0$) with the larger angular momentum number dominates the damping process, since it lasts the longest time (see Tables~\ref{tab:f1}-\ref{tab:f5}).
\begin{table}[!h]
 \tabcolsep 0pt \caption{$\omega$ in $f_{1}(r)$ given by 3-order WKB Approach ($r_{+}=1$,$\beta=\beta_{e}/4$)}\label{tab:f1}
\vspace*{-12pt}
\begin{center}
\def\temptablewidth{1\textwidth}
{\rule{\temptablewidth}{0.5pt}}
\begin{tabular*}{\temptablewidth}{@{\extracolsep{\fill}}ccc}
$m$ &$~~~~~n=0~~~~~$ &$~~~~~n=1~~~~~$ \\
\hline
      $ $  &$|k|=1~~~~~~~~~~~~~~~~~~~|k|=2~~~~~~~~~~~~~~~~~~~|k|=3$&$~~~~|k|=1~~~~~~~~~~~~~~~~~~~|k|=2~~~~~~~~~~~~~~~~~~~|k|=3$ \\
      $0.1$&$0.341-0.205i~~~~~~~0.734-0.185i~~~~~~~1.1146-0.183i$ &$0.321-0.638i~~~~~~~0.696-0.569i~~~~~~~1.086-0.5548i$ \\
      $0.2$&$0.330-0.203i~~~~~~~0.732-0.184i~~~~~~~1.1145-0.182i$ &$0.304-0.618i~~~~~~~0.695-0.568i~~~~~~~1.085-0.5544i$ \\
      $0.3$&$0.321-0.196i~~~~~~~0.731-0.183i~~~~~~~1.1132-0.181i$ &$0.297-0.597i~~~~~~~0.694-0.564i~~~~~~~1.084-0.5533i$ \\
       \end{tabular*}
       {\rule{\temptablewidth}{0.5pt}}
       \end{center}
\tabcolsep 0pt \caption{$\omega$ in $f_{2}(r)$ given by 3-order WKB Approach ($r_{+}=1$,$\beta=\beta_{e}/4$)}\label{tab:f2}
\vspace*{-12pt}
\begin{center}
\def\temptablewidth{1\textwidth}
{\rule{\temptablewidth}{0.5pt}}
\begin{tabular*}{\temptablewidth}{@{\extracolsep{\fill}}ccc}
$m$ &$~~~~~n=0~~~~~$ &$~~~~~n=1~~~~~$ \\
\hline
      $ $  &$|k|=1~~~~~~~~~~~~~~~~~~~|k|=2~~~~~~~~~~~~~~~~~~~|k|=3$&$~~~~|k|=1~~~~~~~~~~~~~~~~~~~|k|=2~~~~~~~~~~~~~~~~~~~|k|=3$ \\
      $0.1$&$0.342-0.204i~~~~~~~0.735-0.1849i~~~~~~~1.1150-0.184i$ &$0.322-0.638i~~~~~~~0.696-0.569i~~~~~~~1.083-0.5550i$ \\
      $0.2$&$0.330-0.203i~~~~~~~0.733-0.1832i~~~~~~~1.1144-0.183i$ &$0.304-0.614i~~~~~~~0.694-0.565i~~~~~~~1.082-0.5548i$ \\
      $0.3$&$0.321-0.198i~~~~~~~0.731-0.1830i~~~~~~~1.1135-0.181i$ &$0.298-0.597i~~~~~~~0.693-0.562i~~~~~~~1.080-0.5537i$ \\
       \end{tabular*}
       {\rule{\temptablewidth}{0.5pt}}
       \end{center}
\tabcolsep 0pt \caption{$\omega$ in $f_{3}(r)$ given by 3-order WKB Approach ($r_{+}=1$,$\beta=\beta_{e}/4$)}\label{tab:f3}
\vspace*{-12pt}
\begin{center}
\def\temptablewidth{1\textwidth}
{\rule{\temptablewidth}{0.5pt}}
\begin{tabular*}{\temptablewidth}{@{\extracolsep{\fill}}ccc}
$m$ &$~~~~~n=0~~~~~$ &$~~~~~n=1~~~~~$ \\
\hline
      $ $  &$|k|=1~~~~~~~~~~~~~~~~~~~|k|=2~~~~~~~~~~~~~~~~~~~|k|=3$&$~~~~|k|=1~~~~~~~~~~~~~~~~~~~|k|=2~~~~~~~~~~~~~~~~~~~|k|=3$ \\
      $0.1$&$0.337-0.208i~~~~~~~0.738-0.185i~~~~~~~1.113-0.184i$ &$0.324-0.680i~~~~~~~0.716-0.607i~~~~~~~1.087-0.5519i$ \\
      $0.2$&$0.329-0.206i~~~~~~~0.736-0.182i~~~~~~~1.112-0.183i$ &$0.310-0.659i~~~~~~~0.714-0.605i~~~~~~~1.086-0.5515i$ \\
      $0.3$&$0.325-0.202i~~~~~~~0.733-0.181i~~~~~~~1.110-0.182i$ &$0.309-0.635i~~~~~~~0.712-0.602i~~~~~~~1.085-0.5509i$ \\
       \end{tabular*}
       {\rule{\temptablewidth}{0.5pt}}
       \end{center}
\tabcolsep 0pt \caption{$\omega$ in $f_{4}(r)$ given by 3-order WKB Approach ($r_{+}=1$,$\beta=\beta_{e}/4$)}\label{tab:f4}
\vspace*{-12pt}
\begin{center}
\def\temptablewidth{1\textwidth}
{\rule{\temptablewidth}{0.5pt}}
\begin{tabular*}{\temptablewidth}{@{\extracolsep{\fill}}ccc}
$m$ &$~~~~~n=0~~~~~$ &$~~~~~n=1~~~~~$ \\
\hline
      $ $  &$|k|=1~~~~~~~~~~~~~~~~~~~|k|=2~~~~~~~~~~~~~~~~~~~|k|=3$&$~~~~|k|=1~~~~~~~~~~~~~~~~~~~|k|=2~~~~~~~~~~~~~~~~~~~|k|=3$ \\
      $0.1$&$0.352-0.218i~~~~~~~0.764-0.188i~~~~~~~1.112-0.185i$ &$0.325-0.679i~~~~~~~0.716-0.608i~~~~~~~1.089-0.552i$ \\
      $0.2$&$0.342-0.217i~~~~~~~0.756-0.187i~~~~~~~1.111-0.184i$ &$0.311-0.658i~~~~~~~0.714-0.606i~~~~~~~1.088-0.551i$ \\
      $0.3$&$0.332-0.209i~~~~~~~0.756-0.186i~~~~~~~1.110-0.182i$ &$0.309-0.636i~~~~~~~0.713-0.602i~~~~~~~1.086-0.550i$ \\
       \end{tabular*}
       {\rule{\temptablewidth}{0.5pt}}
       \end{center}
 \tabcolsep 0pt \caption{$\omega$ in $f_{5}(r)$ given by 3-order WKB Approach ($r_{+}=1$,$\beta=\beta_{e}/4$))}\label{tab:f5}
 \vspace*{-12pt}
\begin{center}
\def\temptablewidth{1\textwidth}
{\rule{\temptablewidth}{0.5pt}}
\begin{tabular*}{\temptablewidth}{@{\extracolsep{\fill}}ccc}
$m$ &$~~~~~n=0~~~~~$ &$~~~~~n=1~~~~~$ \\
\hline
      $ $  &$|k|=1~~~~~~~~~~~~~~~~~~~|k|=2~~~~~~~~~~~~~~~~~~~|k|=3$&$~~~~|k|=1~~~~~~~~~~~~~~~~~~~|k|=2~~~~~~~~~~~~~~~~~~~|k|=3$ \\
      $0.1$&$0.331-0.202i~~~~~~~0.732-0.1841i~~~~~~~1.112-0.182i$ &$0.329-0.675i~~~~~~~0.722-0.605i~~~~~~~1.088-0.559i$ \\
      $0.2$&$0.329-0.201i~~~~~~~0.730-0.1831i~~~~~~~1.111-0.181i$ &$0.314-0.654i~~~~~~~0.721-0.603i~~~~~~~1.085-0.558i$ \\
      $0.3$&$0.321-0.191i~~~~~~~0.729-0.1830i~~~~~~~1.110-0.180i$ &$0.307-0.633i~~~~~~~0.720-0.598i~~~~~~~1.081-0.557i$ \\
       \end{tabular*}
       {\rule{\temptablewidth}{0.5pt}}
       \end{center}
\end{table}

Finally, we adopt the finite difference method to study the dynamical evolution of the
massive Dirac field in the time domain. According to Eq.(\ref{eqfg}), we can get the equations
\begin{equation}
\label{G}
\omega G^{\pm}=-\frac{f(r)dF^{\pm}}{dr}+\sqrt{f(r)}\frac{k}{r}F^{\pm}+m\sqrt{f(r)}G^{\pm},
\end{equation}
\begin{equation}
\label{F}
\omega F^{\pm}=\frac{f(r)dG^{\pm}}{dr}+\sqrt{f(r)}\frac{k}{r}G^{\pm}-m\sqrt{f(r)}F^{\pm}.
\end{equation}
Multiplying $\omega$ on both sides, the above equations become
\begin{equation}
\label{G2}
\omega^{2} G^{\pm}=-\frac{f(r)d(\omega F^{\pm})}{dr}+\sqrt{f(r)}\frac{k}{r}(\omega F^{\pm})+m\sqrt{f(r)}(\omega G^{\pm}),
\end{equation}
\begin{equation}
\label{F2}
\omega^{2} F^{\pm}=\frac{f(r)d(\omega G^{\pm})}{dr}+\sqrt{f(r)}\frac{k}{r}(\omega G^{\pm})-m\sqrt{f(r)}(\omega F^{\pm}).
\end{equation}
Putting Eq.(\ref{G}) and (\ref{F}) into Eq.(\ref{G2}) and (\ref{F2}), then we can derive the differential equations

\begin{figure}
\centering \subfigure{
\begin{minipage}[t]{0.3\textwidth}
\includegraphics[width=1\textwidth]{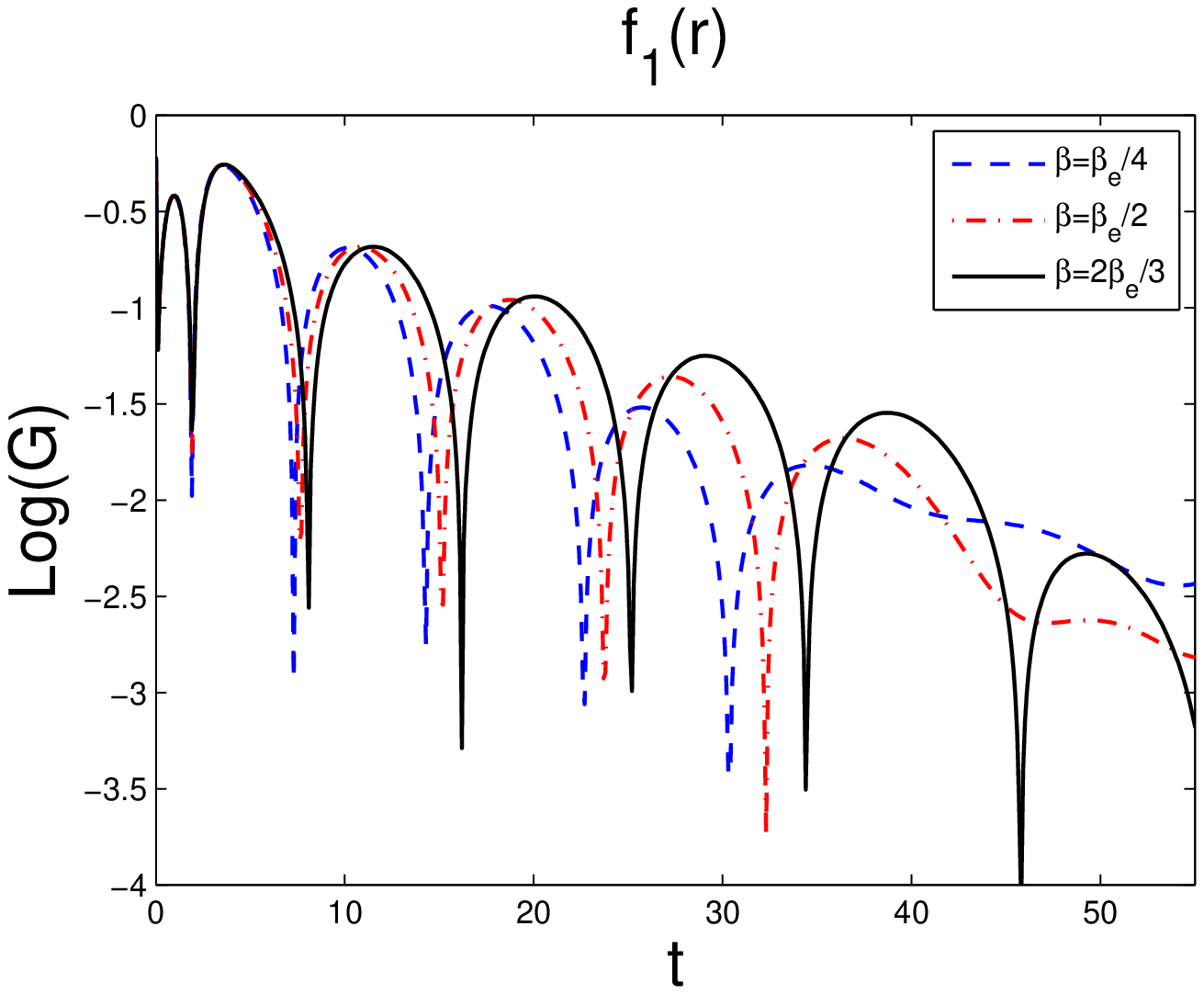} \\
\includegraphics[width=1\textwidth]{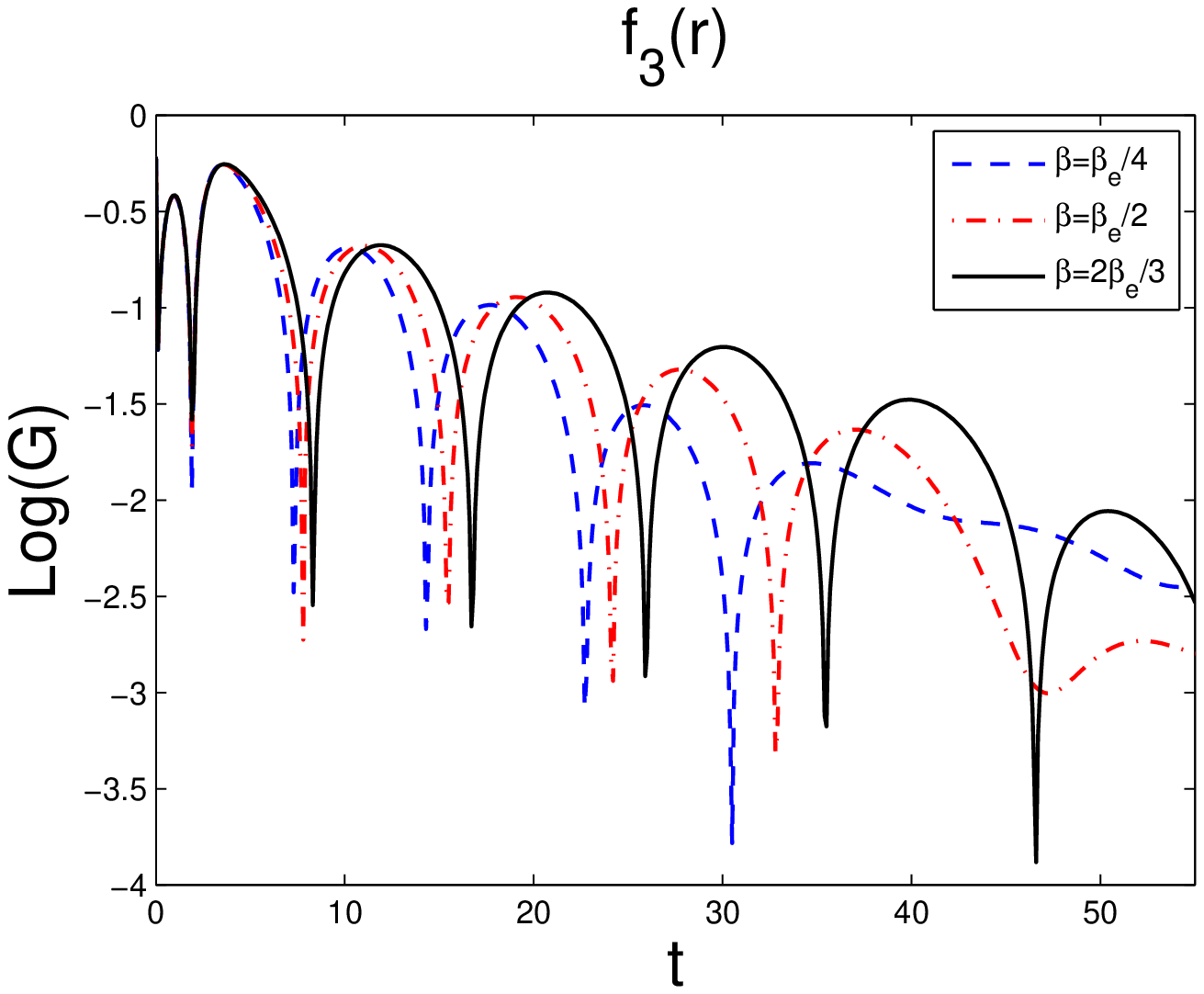} \\
\includegraphics[width=1\textwidth]{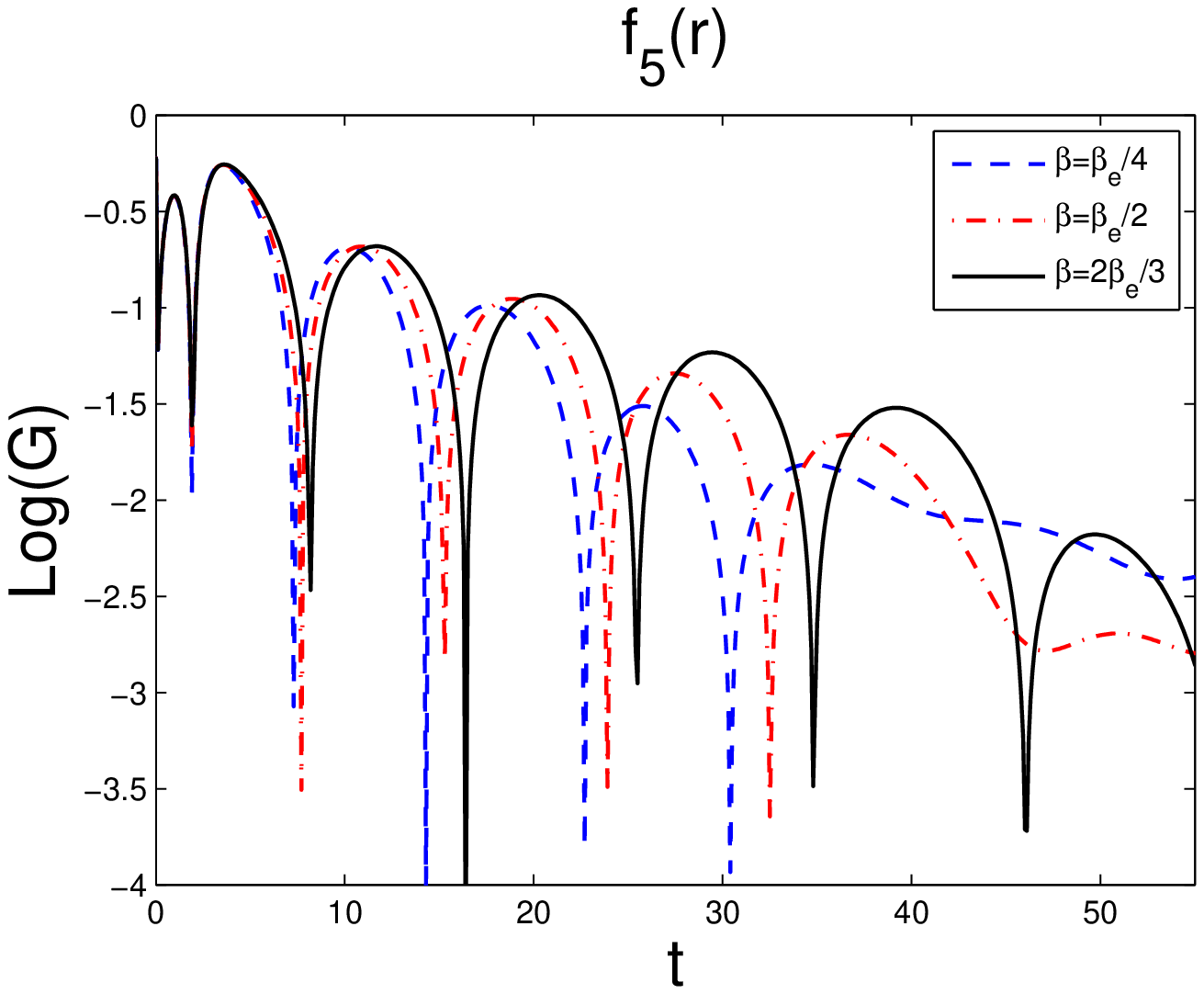}
\end{minipage}
} \subfigure{
\begin{minipage}[t]{0.3\textwidth}
\includegraphics[width=1\textwidth]{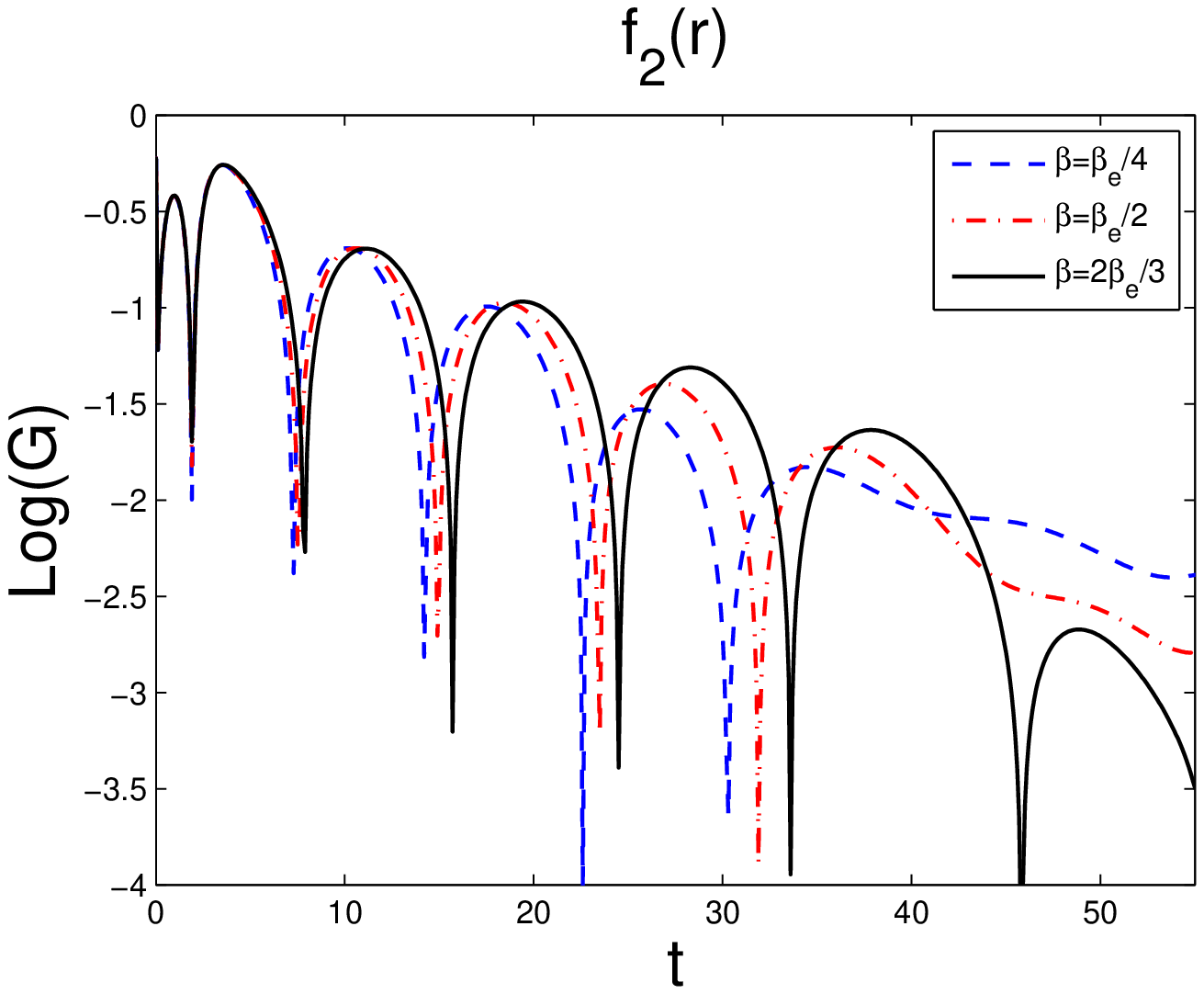} \\
\includegraphics[width=1\textwidth]{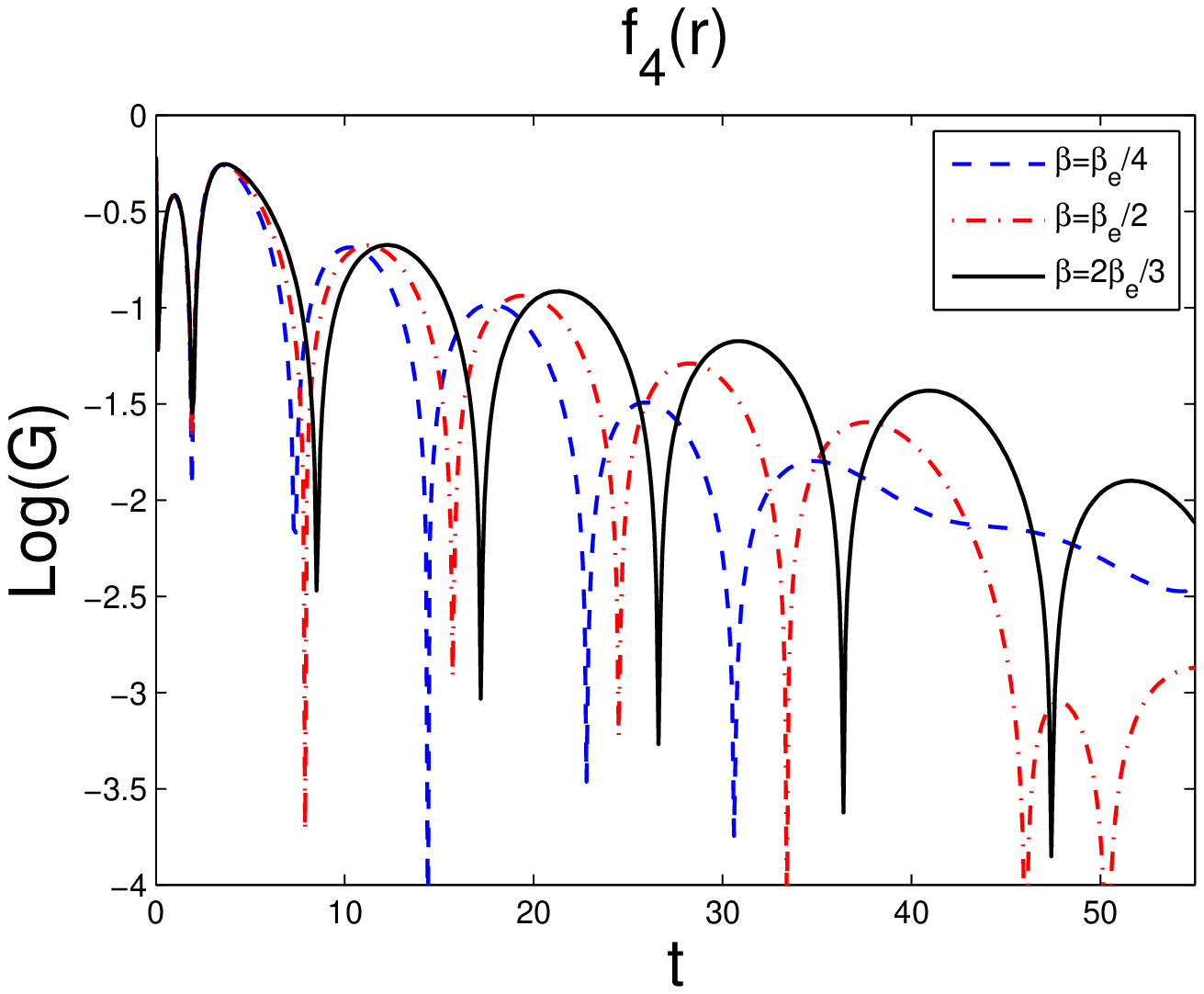}
\end{minipage}
}
\caption{The dynamical evolution of a Dirac field
perturbation to the regular black holes. In each plot, the
dashing, dot-dashing and solid lines correspond to the cases
$\beta=\beta_{e}/4,\beta_{e}/2,2\beta_{e}/3$. We set $r_{+}=1,m=1/4$.}\label{fig:dynamical}
\end{figure}

\bqn\label{FG}
\frac{m\sqrt{f}}{2}f'F^{\pm}&=&f^{2}G''^{\pm}+ff'G'^{\pm}+\left[\frac{kf'}{2r}\sqrt{f}-\frac{k^{2}}{r^{2}}f-
\frac{k}{r^{2}}f^{3/2}+\omega^{2}-m^{2}f\right]G^{\pm};\nb\\
\frac{m\sqrt{f}}{2}f'G^{\pm}&=&f^{2}F''^{\pm}+ff'F'^{\pm}+\left[-\frac{kf'}{2r}\sqrt{f}-\frac{k^{2}}{r^{2}}f+
\frac{k}{r^{2}}f^{3/2}+\omega^{2}-m^{2}f\right]F^{\pm}.\eqn
Here $'$ means $\partial/\partial r$, $\omega^{2}=-\partial^{2}/\partial^{2} t$.
We apply the coordinate transformation $(t,r) \rightarrow (\mu,\nu)$, where $\mu=t-r_{*},\nu=t+r_{*}$, yielding
\bqn
\frac{\partial}{\partial r}=\frac{1}{f}(\frac{\partial}{\partial \nu}-\frac{\partial}{\partial \mu}),~~~~~~
\frac{\partial}{\partial t}=\frac{\partial}{\partial \nu}+\frac{\partial}{\partial \mu}. \eqn
Using this transformation, Eq.(~\ref{FG}) gives differential equations for $F^{\pm}(\mu,\nu)$ and $G^{\pm}(\mu,\nu)$, which can be integrated numerically using the finite difference method suggested in~\cite{Finit1,Finit2,Jin8}. Fig.~\ref{fig:dynamical} shows the detailed evolution of a massive Dirac field in regular BHs with $m=1/4, |k|=1$ and $\beta=\beta_{e}/4,\beta_{e}/2,2\beta_{e}/3$. Over this range of $\beta$, we find that the damping speed becomes slower with increasing $\beta$.
This is in accordance with the massless case.

\section{conclusion}\label{sec:conclusions}

We have studied several spherically symmetric regular spacetimes. All of them reduce to a Schwarzschild black hole when $\beta=0$. For non-zero $\beta$, they have no singularity---even at $r=0$, $f(r)$ is finite. Furthermore, there is more than one horizon for $\beta \ne 0$, which may lead to extreme black holes for $\beta > \beta_e$. The horizon obtained for an extreme charge reflects the different horizon locations for each of the different spherically symmetric regular black holes we have studied. The calculation of extreme charge can be extended to other black hole spacetimes, but in the case of non-spherically symmetric BHs (e.g. Kerr) the situation is complicated by the fact that there is no static expression for the horizons.

In the massless case, the effective potentials and QNMs for all the regular spacetimes have similar behavior based on the angular momentum number $k$ and the magnetic monopole charge $\beta$. The potentials increase with increasing $|k|$ so that the oscillation becomes more intense while the decay proceeds more slowly. In addition, the existence of the magnetic monopole charge $\beta$ causes the QNM frequencies to differ from the Schwarzschild case, with longer damping times.

In the massive case, the effective potential follows the same asymptotic behavior as the Schwarzschild case ($V(r \rightarrow \infty) = m^2$). The decay rate of the BH perturbation decreases with increasing mass $m$ of the field. As one of the highlights of this article, we decouple the Dirac wave equation and derive the analytic relationship between the mass of the field and the frequency of the QNM. This work can be applied to other massive field perturbations.

As one specific case of Dirac perturbation, above results also illuminate the role of spin on the frequency spectrum. Comparing with Dirac-Schwarzschild BHs of Cho~\cite{Diracfield,q3jia1}, the QNMs frequencies occur similar behavior in massless and massive cases. On the other hand, the scalar perturbation to Schwarzschild~\cite{RegularBH,Massivescalar} and Bardeen BHs~\cite{RegularQuasinormal1} cause the similar behavior of damping rate ($\text{Im}(\omega)$) as for the cases in this paper, but lead to different behavior of oscillation frequency ($\text{Re}(\omega)$), which show the charge $\beta$ would increase the frequency of oscillation and the mass of field $m$ enhance the lowest oscillation frequency. Furthermore for the first overtone, the $\text{Re}(\omega)$ reaches a maximum and decrease rapidly with $m$. Those indicate we can distinguish the spin of field around BHs from its QNMs.

There are a lot of problems left for future study. Firstly, according to the QN ringing (see Fig.3) the difference between RN (singular black holes) and regular BHs is subtle, which is only prominent for relatively large values of
the charge (with respect to the critical value). Thus, the difference is only observable in principle. As the QNM of BHs rely on what
happens outside the horizon, it would be difficult to distinguish between regular and singular black holes through measuring the QN ringing. Secondly, the regular space time we discussed is spherically symmetric. In general, the study of
non-spherically symmetric cases (such as Kerr black hole) would give us much more
useful information. Thirdly there are still many useful regular BHs not discussed in this paper, such as the regular solution with de Sitter center instead of a singularity presented in ~\cite{new1}; Another kind of regular BHs called "cold black holes" with horizons of infinite area having no center at all ~\cite{new2,new3}; In addition, some regular BHs without a center, with an expanding, asymptotically
de Sitter universe inside the horizon called as "black universes" ~\cite{new4,new5,new6}. Therefore we plan to study they in next work. Finally, how to
combine the knowledge of quantum gravity to QNMs frequencies is
still un-known.
\section*{\bf Acknowledgements}
Our heartfelt thanks to Professor Matthew Benacquista for great helpful discussion. Meanwhile we want to express our appreciation to professor Bronnikov, Alexander Zhidenko, Roman Konoplya, Dymnikova and Ayon-Beato for their useful suggestion. This work was supported by FAPESP No. 2012/08934-0, National Natural Science Foundation of China No. 11205254, No. 11178018 and No.
11075224, and the Natural Science Foundation Project of CQ CSTC
2011BB0052, and the Fundamental Research Funds for the Central Universities (project No. CDJRC10300003).


\end{document}